\documentclass[11pt,a4paper]{article}
\pdfoutput=1  
\usepackage{epsfig}
\usepackage{graphicx,psfrag}
\usepackage{multirow}
\usepackage{slashed}
\usepackage{pstricks}
\usepackage{caption}
\usepackage{subcaption}
\usepackage{url}
\usepackage{braket}
\usepackage{jheppub}
\usepackage{enumerate}
\usepackage{wasysym} 
\usepackage{mathrsfs} 
\usepackage{amsfonts} 
\usepackage{amsbsy} 
\usepackage{amscd}
\usepackage{color}
\usepackage{changepage}

\makeatletter

\def\eeq{\end{equation}}
\def\beq{\begin{equation}}

\newcommand{\Rmnum}[1]{\expandafter\@slowromancap\romannumeral #1@}

\newcommand{\gsim}{\raisebox{-0.13cm}{~\shortstack{$>$ \\[-0.07cm]
      $\sim$}}~}
\newcommand{\lsim}{\raisebox{-0.13cm}{~\shortstack{$<$ \\[-0.07cm]
      $\sim$}}~}
\makeatother

\title{The mono-Higgs + MET signal at the Large Hadron Collider: a study on the $\gamma\gamma$ and $b\bar{b}$ final states}

\author[a]{Debabrata Bhowmik,}
  \affiliation[a]{High Energy Nuclear and Particle Physics Division, Saha Institute of Nuclear Physics,HBNI,
 1/AF Bidhannagar, Kolkata - 700064, India }

\author[b]{Jayita Lahiri,} 
  \affiliation[b]{Regional Centre for Accelerator-based Particle Physics,
Harish-Chandra Research Institute, HBNI,
Chhatnag Road, Jhunsi, Allahabad - 211 019, India} 

\author[a]{Satyaki Bhattacharya,}
  \affiliation[a]{High Energy Nuclear and Particle Physics Division, Saha Institute of Nuclear Physics,HBNI,
 1/AF Bidhannagar, Kolkata - 700064, India }

\author[c]{Biswarup Mukhopadhyaya,} 
\author[c]{Ritesh K Singh} 
  \affiliation[c]{Department of Physical Sciences, Indian Institute of Science Education and Research Kolkata, Mohanpur - 741246, India}

\abstract{
We investigate the potential of the channel {\em mono-Higgs + MET} in yielding signals
of dark mater at the high-luminosity Large Hadron Collider (LHC). As illustration,
a scalar dark matter in a Higgs portal scenario has been chosen, whose phenomenological
viability has been ensured by postulating the existence of dimension-6 operators
that enable cancellation in certain amplitudes for elastic scattering of dark matter
in direct search experiments. These operators are found to have non-negligible contribution
to the mono-Higgs signal. Thereafter, we carry out a detailed analysis of this 
signal, with the accompanying MET providing a useful handle in suppressing
backgrounds. Signals for the Higgs decaying into both the diphoton  and $b{\bar b}$ 
channels have been studied. A cut-based
simulation is presented first, followed by a demonstration of how the statistical
significance can be improved through analyses based on Boosted Decision Trees
and Artificial Neural Network. The improvement is found to be especially noticeable
for the $b{\bar b}$ channel.

}

\preprint{HRI-RECAPP-2020-011\\$\textrm{}$}

\begin{document}

\maketitle

\newpage

\section{Introduction}

The Standard Model(SM) of particle physics has proven to be an
extremely successful theory so far. Experimental studies have
confirmed most of its predictions to impressive levels of accuracy\cite{Abe:1995hr,D0:1995jca,Drees:2001xw,ALEPH:2005ab,Khachatryan:2016vau}.
It still remains an intense quest to look for physics beyond the
standard model. Perhaps the most concrete and persistent reason for
this is the existence of dark matter (DM) which constitutes to upto
23\% of the energy density of the universe, and the belief that DM
owes its origin to some hitherto unseen elementary particle(s). In such a
situation, one would like to know if the DM particle interacts with
those in the SM, and if so, what the signatures of such interactions
will be.  The literature is replete with ideas as to the nature of DM,
one historically curious possibility being one or more weakly
interacting massive particle(s) (WIMP), with the DM particle(s)
interacting with those in SM particles  coupling strength of the order of
the weak interaction strength.

The collider signal of a WIMP DM is naturally expected to consist in
missing-$E_T$ (MET) in association with visible particle(s) that can
be easily tagged. Final states such as monojet, mono-photon, monos-$V$ etc. are
frequently explored in this spirit~\cite{Khachatryan:2014rra,Diehl:2014dda,No:2015xqa,Schramm:2016csx,Aaboud:2016uro,ATLAS:2018jsf,CMS:2018fux}. 
It may be asked whether one can
similarly have mono-Higgs DM signals, accompanied by hard MET caused
by DM pairs (assuming that a $Z_2$ symmetry makes the DM stable)~\cite{Ghorbani:2016edw,Miniello:2017esf}. Such analyses in the context of supersymmetric models have been performed in the past~\cite{Abdallah:2016vcn,Baum:2017gbj}. The
existing studies in this context leave enough scope for refinement,
including (a) thorough analyses of the proposed signals as well as
their SM backgrounds at the Large Hadron Collider (LHC), and (b) the
viability of {\it Higgs + DM-pair} production, consistently with
already available direct search constraints \cite{Carpenter:2013xra,Petrov:2013nia,Berlin:2014cfa,Basso:2015aee}.  Such
constraints already disfavour so-called `Higgs portal' scenarios in
their simplest versions\cite{Djouadi:2012zc,Greljo:2013wja,Han:2016gyy,Aaboud:2019yqu,Arcadi:2019lka}. However, there are foreseeable
theoretical proposals\cite{Gross:2017dan,Dey:2019lyr,Okada:2020zxo} involving new physics, where the
Higgs-mediated contribution to spin-independent cross-section in
direct search experiments undergo cancellations from additional
contributing agents. Keeping this in mind, as also the fact that LHC
is not far from its high-luminosity phase, it is desirable to sharpen
search strategies for mono-Higgs + MET signals anyway, especially because it
relates to the appealing idea that the Higgs sector is the gateway to
new physics.  However, such signals are understandably
background-prone, and refinement of the predictions in a realistic LHC
environment is a necessity. Furthermore, it needs to be ascertained
how the additional terms in the low-energy theory cancelling the Higgs
contributions in direct search experiments affect searches at the
LHC. We address both issues in the current study.

As for the additional terms cancelling the contributions of the
125-GeV scalar to spin-independent inelastic scattering,
scenarios with an extended Higgs sector have been studied
earlier~\cite{Dey:2019lyr}. Here, however, we take a
model-independent approach, and postulate the new physics effects to
come from dimension-6 and-8 operators which are suppressed by the
scale of new physics. These lead to the rather interesting possibility
of partial cancellation between the coefficients of dimension-4 and
higher dimension operators. Thus at the same time, one obeys direct
detection constraints, has not-so-small coupling between the Higgs
and the DM, and matches the observed relic density. And it is the higher-dimensional
operators that play significant roles in mono-Higgs production at the LHC.

We consider for illustration the $\gamma \gamma$ and $b {\bar b}$ decay modes
of the mono-Higgs, along with substantial MET. 
We have started with  rectangular cut-based analyses for both final states.
The di-photon events  not only have the usual SM backgrounds but also can be faked
to a substantial degree by $e\gamma$-enriched di-jet events.  Following up
on a cut-based analysis, we switch on to machine learning techniques to improve the
signal significance, going all the way to using
artificial neural networks(ANN) and boosted decision tree(BDT) for both $\gamma \gamma$ and $b \bar b$ final states.

The plan of our paper is as follows. In Section~\ref{sec2} we discuss the outline of the model-independent scenario that we have considered and we take into account all the relevant constraints on this scenario and find out viable and interesting parameter space which can give rise to substantial mono-Higgs signature at the high luminosity LHC. In Section~\ref{sec4} we discuss in detail our signals and all the major background processes. In Section~\ref{sec5} we present our results of a rectangular cut-based analysis. In Section~\ref{sec6} we employ machine-learning tools to gain improved signal significance over our cut-based analysis. In Section~\ref{sec7} we summarize our results and conclude the discussion.

\section{Outline of the scenario and its constraints}\label{sec2}

\subsection{The theoretical scenario}
We illustrate our main results in the context of a scenario of a scalar DM particle.
Where scalar sector is augmented by the gauge-singlet $\chi$, the potential can be generally written by

\begin{equation} 
{\cal V} = {\cal V}(\Phi) + {\lambda}_{\Phi \chi} \Phi^{\dagger} \Phi \chi^2 + \lambda_{\chi} \chi^4 
\end{equation}

Here $\Phi$ is the SM Higgs doublet and ${\lambda}_{\Phi \chi}$ and $\lambda_{\chi}$ are the relevant quartic couplings. 


\begin{figure}[!hptb]
\centering
\includegraphics[width=6.5cm, height=4.5cm]{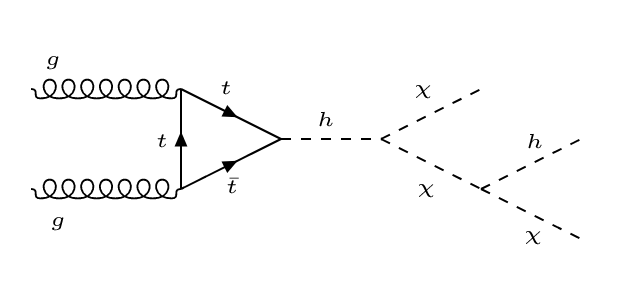} 
\includegraphics[width=6.5cm, height=4.0cm]{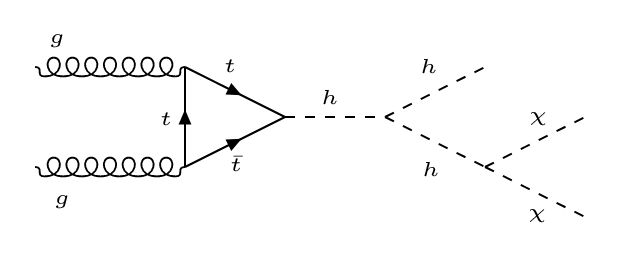} \\
\includegraphics[width=6.0cm, height=4cm]{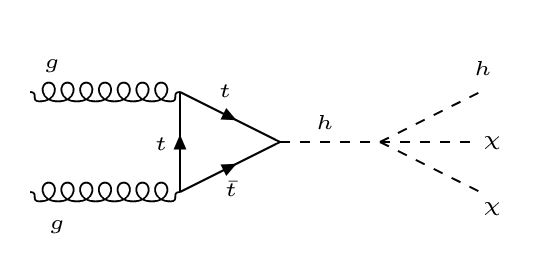}
\caption{The contribution to mono-Higgs + $\slashed{E_T}$ final state from dimension -4 operators.}
\label{feynhportal}
\end{figure}
\begin{figure}[!hptb]
\centering
\includegraphics[width=6.5cm, height=4.5cm]{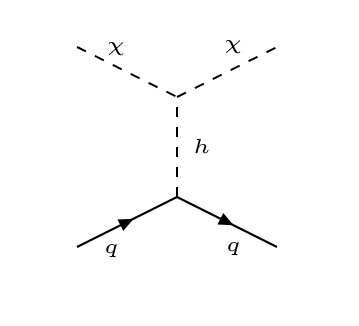} 
\includegraphics[width=6.5cm, height=4.5cm]{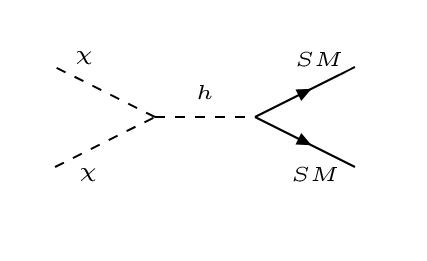}
\caption{The contribution to $\chi$-nucleon elastic scattering cross-section (left) and observed relic density of the universe (right) from the dimension-4 operator.}
\label{ddrelic}
\end{figure}

\noindent
It is clear from above that the simplest operator involving the Higgs boson and a scalar DM $\chi$
is the dimension-4 renormalizable operator $\Phi^{\dagger} \Phi
\chi^2$.  This operator gives rise to
the dominant contribution to the mono-Higgs + $\slashed{E_T}$ final state
when Higgs is produced via gluon fusion, as can be seen from
Figure~\ref{feynhportal}.  This operator also takes part in the
DM-nucleon elastic scattering through $t$-channel Higgs exchange (see
Figure~\ref{ddrelic} (left)) and DM annihilation diagram with
$s$-channel Higgs mediation (Figure~\ref{ddrelic} (right)). The stand-alone presence
of this operator makes it difficult to satisfy both direct detection
constraints and relic density requirements simultaneously, as will be
discussed in the next section. However, one can go beyond dimension-4
terms and construct higher-dimensional operators involving
(anti)quarks, Higgs and a pair of DM particles, which encapsulate the
entire contribution to the mono-Higgs + $\slashed{E_T}$ signal, and at
the same time has the potential to cancel the contribution of the
aforementioned dimension-4 operator to the spin-independent
cross-section.

One can write two SU(2)$_L \times$ U(1)$_Y$ gauge-invariant and
Lorentz invariant operators in this context, namely ${\cal O}_1$ and
${\cal O}_2$, which are of dimension-6 and-8 respectively, as follows:

\begin{eqnarray}\label{eqnscalar}
{\cal O}_1 &=& \frac{1}{\Lambda^2} (\bar{Q}_L \Phi d_R \chi^2 +
\bar{Q}_L \tilde{\Phi} u_R \chi^2) \\ {\cal O}_2 &=&
\frac{1}{\Lambda^4} (\bar{Q}_L D^{\mu}\Phi d_R \chi \partial_{\mu}\chi
+ \bar{Q}_L D^{\mu}\tilde{\Phi} u_R \chi \partial_{\mu}\chi)
\end{eqnarray}

\noindent
${\cal O}_1$ is a dimension-6 operator involving a quark-antiquark
pair, the Higgs boson and a pair of DM particles. On the other hand,
${\cal O}_2$ is of dimension-8, involving derivatives of
$\Phi$ as well as $\chi$.  Both of these operators are
suppressed by a high-scale $\Lambda$ where some unknown new physics
is believed to exist.

\begin{figure}[!hptb]
\includegraphics[width=6.3cm, height=5cm]{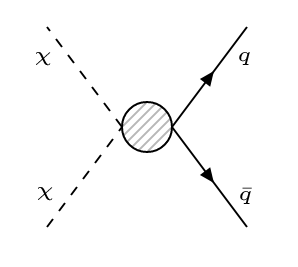}
\hspace{0.00001cm}
\includegraphics[width=6.3cm, height=5cm]{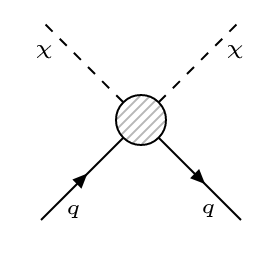}
\centering
\caption{Generic diagrams for annihilation of DM pair into SM states~(left) and DM-nucleon elastic scattering (right) with contributions from dimension-6 and -8 operators.}
\label{fig3}
\end{figure}

Once the higher dimensional operators are introduced, they contribute to both spin-independent cross-section in direct searches and annihilation of $\chi$ before freeze-out. The appropriate feynman diagrams are shown in Figure~\ref{fig3}. One should note that contributions in Figure~\ref{fig3} arises when $\Phi$ acquires a VEV. Such contributions will interfere with those coming from the diagrams shown in Figure~\ref{ddrelic}. 

\begin{figure}[!hptb]
\centering
\includegraphics[width=6.5cm, height=5.5cm]{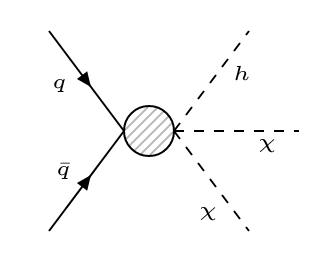} 
\caption{Generic diagram for mono-Higgs + $\slashed{E_T}$ signal with contributions from dimension-6 and -8 operators.}
\label{pphchichi}
\end{figure}

Let us also mention the operators ${\cal O}_1$ and ${\cal O}_2$ contribute only to the {\it spin-independent} cross-section in direct search due to the absence of $\gamma_5$ in them.
And finally, the presence of higher dimensional operators also opens up additional production channels leading to the mono-Higgs signals via quark induced diagrams, whose generic representation can be found in Figure~\ref{pphchichi}.

\subsection{Constraints from the dark matter sector and allowed parameter space}

As the scenario under consideration treats $\chi$ as a weakly interacting thermal dark matter candidate, 
it should satisfy the following constraints:
\begin{itemize}
\item The thermal relic density of $\chi$ should be consistent with the latest Planck limits
at the 95\% confidence level~\cite{Ade:2013zuv}. 

\item The $\chi$-nucleon cross-section should be below 
the upper bound given by XENON1T experiment~\cite{Aprile:2018dbl} and any other
data as and when they come up.

\item Indirect detection constraints coming from both isotropic gamma-ray data
and the gamma ray observations from dwarf spheroidal galaxies~\cite{Ackermann:2015zua} 
should be satisfied at the 95\% confidence level. This in turn
puts an upper limit on the velocity-averaged $\chi$-annihilation cross-section~\cite{Ahnen:2016qkx}.

\item The invisible decay of the 125-GeV scalar Higgs $h$ has to be $\leq$
19\%~\cite{Sirunyan:2018owy}. 
\end{itemize}

\begin{figure}[!hptb]
\centering
\includegraphics[width=12.0cm, height=10cm]{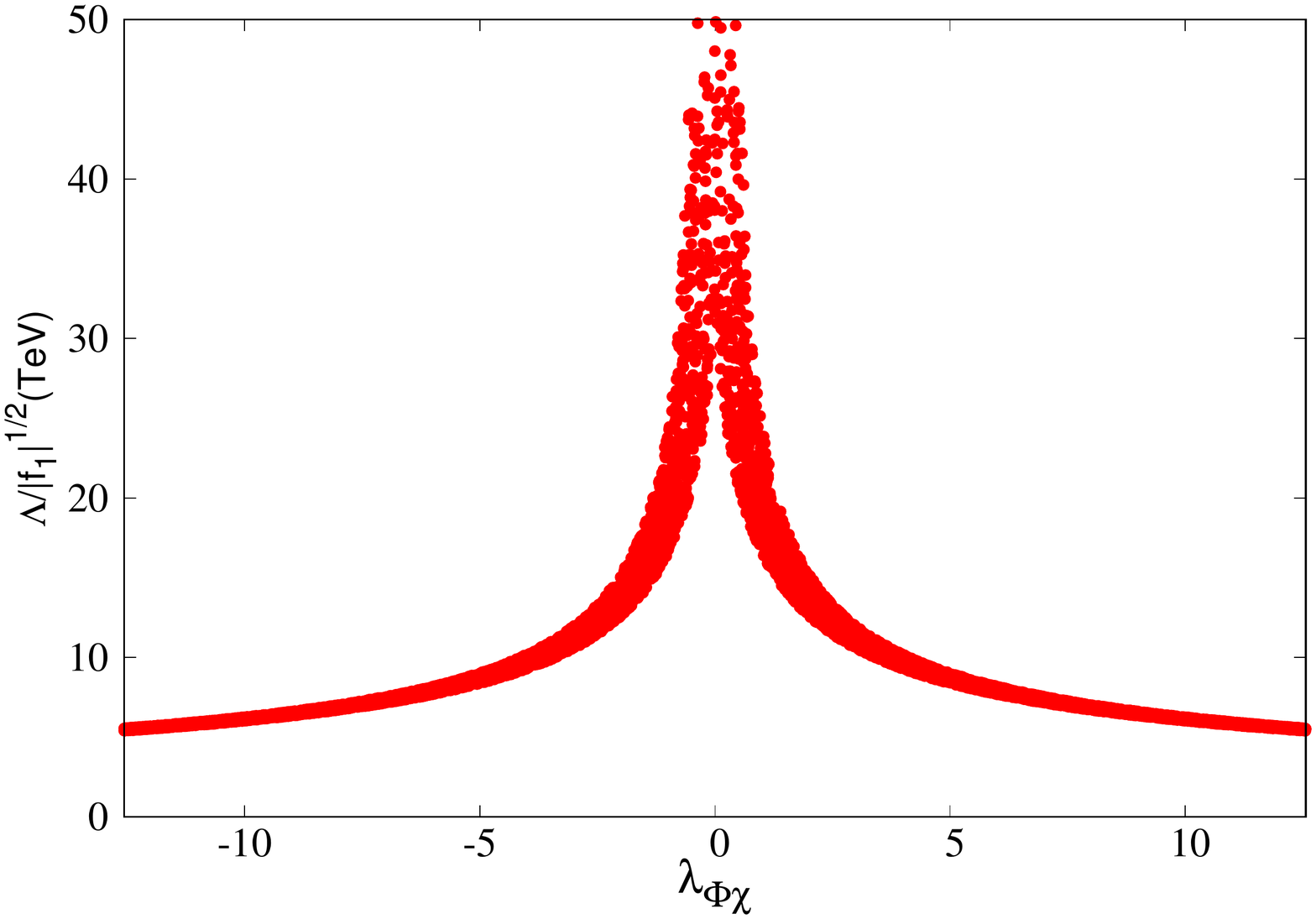} 
\caption{Region allowed by the relic density and direct detection observation in the parameter space spanned by the coefficients of dimension-4 and dimension-6 operators for scalar DM.}
\label{lam2_lam3}
\end{figure}

It has been already mentioned that the simplest models using the SM
Higgs as the dark matter portal is subject to severe constraints. The
constraints are two-fold: from the direct search results, especially
those from Xenon-1T~\cite{Aprile:2018dbl}, and from the estimates of relic density, the
most recent one coming from Planck~\cite{Ade:2013zuv}. While the simultaneous
satisfaction of both constraints restricts SM Higgs portal scenarios
rather strongly, the same restrictions apply to additional terms in
the Lagrangian as well. In our case, the coefficient of $\Phi^\dagger \Phi \chi^2$,
restricted to be ultra-small from direct search data, cannot ensure the requisite 
annihilation  rate.

\begin{figure}[!hptb]
\centering
\includegraphics[width=12.0cm, height=10cm]{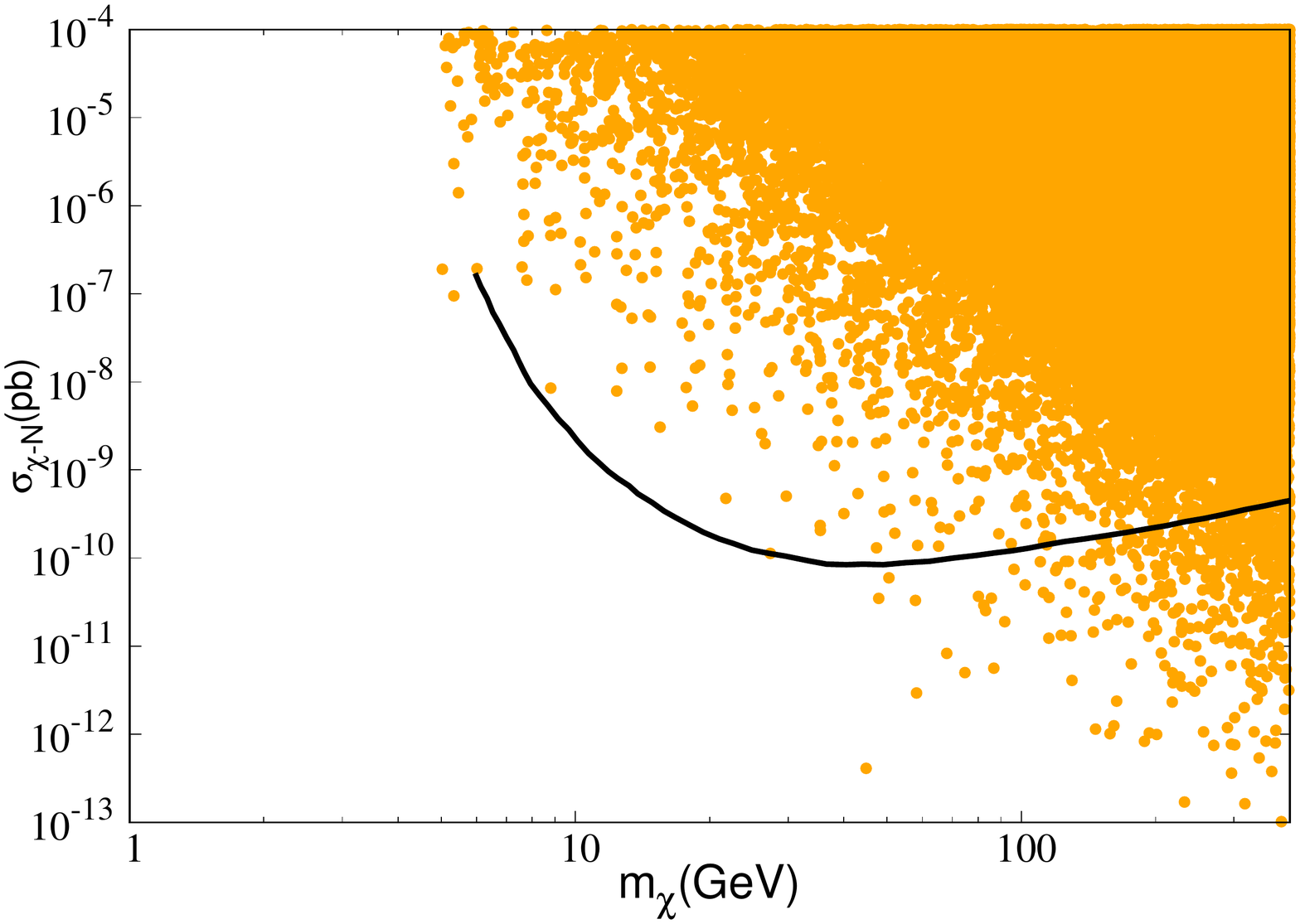} 
\caption{Parameter space (yellow points) allowed by the relic density observation. The black line is the upper limit on the spin-independent $\chi-N$ scattering cross-section from XENON1T experiment. Region below the black line is allowed by direct detection bound from the XENON1T experiment.}
\label{sigma_dd}
\end{figure}

It is thus imperative to have additional terms that might 
bring out cancellation of the Higgs contribution in direct search and thus 
make the quartic term less constrained.
The signs of the tri-linear coupling $\lambda$ and the Wilson coefficients $f_i$s 
have to be appropriately positive or negative to ensure destructive interference. 
While such cancellation may apparently be inexplicable,
it is important to phenomenologically examine its implication, in a model-independent
approach if possible. A similar approach has been taken in a number of recent works~\cite{Gross:2017dan,Dey:2019lyr,Okada:2020zxo}. The higher-dimensional operators listed in the previous subsection
are introduced in this spirit. We must add that such cancellation is conceivable so as $\frac{y_s}{m_h^2}$ is comparable to $\frac{|f_1|}{\Lambda^2}$ where $y_s$ and $f_1$ are respectively the SM strange quark Yukawa coupling and the Wilson coefficient of the effective operator in ${\cal O}_1$.



In Figure~\ref{lam2_lam3}, we show the
region of parameter space in the $\frac{1}{v}\lambda_{h\chi\chi} -
\frac{\Lambda}{\sqrt{|f_1|}}$ plane (considering the ${\cal O}_1$ for illustration), which is consistent with both of the aforesaid
constraints. The broadness of red line owes itself to the $2 \sigma $ band of the relic density as well as the the direct detection limit. As the trilinear coupling decreases, the high scale gets
pushed to a higher value, as expected. A similar cancellation takes
place in the annihilation of DM particles into a pair of quarks as
well. However, in such a situation, too, DM pair annihilating into 
pairs of gauge or Higgs bosons helps us achieve the required 
annihilation cross-section. The red bands in Figure~\ref{lam2_lam3} satisfy both direct detection and relic density constraints.

In Figure~\ref{sigma_dd}, we show regions of the parameter space consistent with the observed relic density (yellow points) as a function of dark matter mass $m_{\chi}$. Here also we have taken the contribution from ${\cal O}_1$ for simplicity.\footnote{The spin-independent cross-section in direct search, however receives considerably lower contributions from ${\cal O}_2$, because of velocity suppression for a non-relativistic DM candidate. We have included both the contributions from ${\cal O}_1$ and ${\cal O}_2$ in mono-Higgs production at the LHC. However the estimates pertain to $f_1 = f_2$ which may not be valid in some theoretical scenarios.}

The black line in the figure represents the upper limit from Xenon-1T on
the spin-independent DM-nucleon elastic scattering cross-section as a function of the mass
of the DM particle. The region below this curve is our allowed parameter space.

\section{Signals and backgrounds}\label{sec4}

Having identified the regions of allowed parameter space we proceed towards developing strategies to probe such scenarios at the high luminosity LHC. Our study is based on a scalar DM $\chi$ as mentioned earlier. One should note that a corresponding fermionic DM will not allow the production channels in Figure~\ref{feynhportal} purely driven by dimension-4 operator. Therefore one will have to depend on higher-dimensional operators with the production rate considerably suppressed.
As has been discussed earlier, we are looking for the mono-Higgs + $\slashed{E_T}$ final state. Since the process will always involve missing energy because of the presence of the stable DM candidates, the decay products of the Higgs constitute the visible system recoiling against the missing transverse momenta. The main contribution to production comes from the process depicted in Figure~\ref{feynhportal}(top left).

\begin{figure}[!hptb]
\centering
\includegraphics[width=12.0cm, height=10cm]{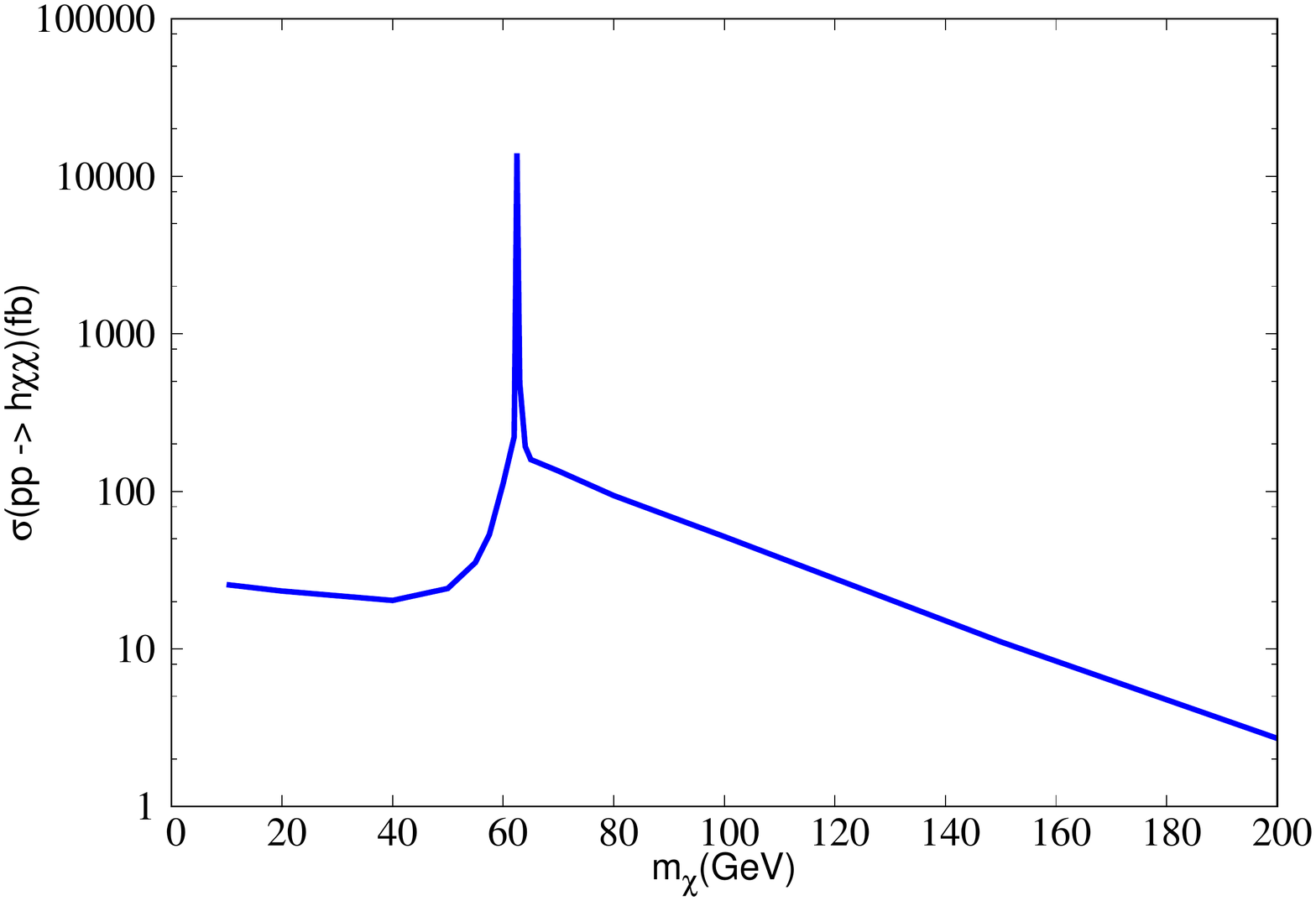} 
\caption{Production cross-section of $p p \rightarrow h \chi \chi$ at 13 TeV as a function of dark matter mass, $\lambda_{\Phi\chi} \approx 4\pi$ and $\Lambda \approx 5$ TeV.  }
\label{crosssec}
\end{figure}

In Figure~\ref{crosssec}, we show the dependence of $\sigma(p p \rightarrow h \chi \chi)$ on $m_{\chi}$ for $\lambda_{\Phi\chi} \approx 4\pi$ and $\frac{\Lambda}{\sqrt{|f_{1,2}|}} \approx 5$ TeV. It is clear from this figure that a resonance takes place in the vicinity of $\frac{m_h}{2}$. 
It is worth mentioning that the effective operator ${\cal O}_1$ contributes close to 10\% as much as the gluon fusion channel in $h \chi \chi$ production, the contribution of ${\cal O}_2$ is about half of that of ${\cal O}_1$. Here also the assumption $f_1 \approx f_2$ is made.

The next important task is to identify suitable visible final states which will recoil against the invisible $\chi \chi$ system. The largest branching ratio of the 125 GeV scalar is seen in the $b \bar b$ channel. However, while this assures one of a copious event rate, one is also
deterred by the very large QCD backgrounds, whose tail poses a threat to the signal significance. While we keep the $b \bar b$ channel within the purview of this study, we start with a relatively cleaner final state, namely a di-photon pair. We, however, are conscious of the corresponding disadvantage, due to the relatively small branching ratio ($\approx 2.27 \times 10^{-3}$). We suggest in the discussion below, some strategies to overcome this disadvantage largely making use of one feature of the signal, namely a substantial missing $\slashed{E_T}$ generated by the $\chi \chi$ system.  


\subsection{$\gamma \gamma +\slashed{E_T}$ channel}

The di-photon channel is apparently one of the cleanest
of Higgs signals. The absence of hadronic products is perceived as 
the main source of its cleanliness, together with the fact that there is 
a branching ratio suppression (though rather strong) at a single level only as opposed to the four-lepton final state.
While this channel has been under scrutiny from the earliest days of
Higgs-related studies at the LHC, in the present context we are focussing on
those events where the two photons, each  with $p_T > 30$ GeV,
recoil against an invisible component which lends a substantial
$\slashed{E_T}$ to the events. The photons are subjected to the isolation criterion
defined by ${\frac{|p^{\gamma}_T|}{\Sigma^i |p^i_T|}} < 0.12$, where the summation in 
the denominator is over all the particles within $\Delta R = 0.5$ around each candidate photon.
Searches for such events have been carried out by both CMS~\cite{CMS:2016xok,CMS-PAS-EXO-16-054,Sirunyan:2017hnk,Sirunyan:2018fpy,Sirunyan:2019zav} and ATLAS~\cite{Aad:2015yga,ATLAS-CONF-2017-024}.  

As can be seen in Figure~\ref{crosssec}, this channel is usable for $m_{\chi} \lsim 100$ GeV,
and particularly in the resonant region, due to the rather small branching ratio
for  Higgs decay into di-photons. Moreover, the upper limit on the invisible
decay of the Higgs prompts us to those benchmarks where $m_{\chi} > m_h/2$.
A set of such benchmark points, satisfying also all constraints related to
dark matter, are listed in Table~\ref{bpgam}. BP2 here has been chosen to be the best possible scenario, with mass of the dark matter in resonant region and large $\lambda_{\Phi \chi}$ coupling (satisfying the perturbativity limit), resulting in largest possible production rate. In case of BP1 we probe the reach of the $\gamma \gamma +\slashed{E_T}$ channel in terms of the dark matter mass with $\lambda_{\Phi \chi}$ still around the perturbative limit. In BP3 we intend to probe the reach of the aforementioned channel in terms of the quartic coupling $\lambda_{\Phi \chi}$. Therefore here we have taken the mass of the dark matter in the resonant region and $\lambda_{\Phi \chi}$ to be lower than that of BP1 and BP2.    

The apparent cleanliness of the signal, however, can be misleading.
Various backgrounds as well as possibilities of misidentification
or mismeasurement tend to vitiate the signal. In order to meet such challenges,
the first step is to understand the backgrounds.

\begin{table}[!hptb]
\begin{center}
\begin{footnotesize}
\begin{tabular}{| c | c | c | c |}
\hline
Benchmarks & $m_{\chi}$ & $\lambda_{\Phi \chi}$ & $\Lambda$ \\
\hline
BP1 &70.0 & 12.0  & 5 TeV \\
\hline
BP2 & 64.0 & 12.0 & 5 TeV \\
\hline
BP3 & 64.0 & 9.0 & 6 TeV \\
\hline
\end{tabular}
\end{footnotesize}
\caption{The benchmark points for $\gamma \gamma + \slashed{E_T}$ final state.}
\label{bpgam}
\end{center}
\end{table}

\noindent
{\bf Backgrounds}: Contamination to the di-photon final state comes mainly from prompt photons that originate from the hard scattering process of the partonic system (e.g. $q\bar{q} \rightarrow \gamma \gamma$ through Born process or $gg \rightarrow \gamma \gamma$ through a one-loop process represented by ``box diagram'') or non-prompt photons, that originate within a hadronic jet, either from hadrons that decay to photons or are created in the process of fragmentation, governed through the quark to photon and gluon to photon fragmentation function $D^q_{\gamma}$ and $D^g_{\gamma}$~\cite{Bourhis:1997yu,Bourhis:2000gs,Kniehl:2000fe,Binnewies:1995kg,Kretzer:2000yf}. Such non-prompt photons are always present in a jet, and can be misidentified as a prompt photon when most of the jet energy is carried by one or more of these photons. We shall refer to this effect as ``jet faking photons". Electrons with energy deposit in the electromagnetic calorimeter(ECAL), can be misidentified as a photon if the track reconstruction process fails to reconstruct the trajectory of the electron in the inner tracking volume, since both electron and photon deposit energy in the ECAL by producing an electromagnetic shower, with very similar energy deposit patterns (shower shapes). Therefore processes with energetic electrons in the final state can also contribute to the background. We shall refer to this type of misidentification as ``electron faking photon". It may be noted that the ``jet faking photons " also has a contribution from non-prompt electrons produced inside the jet, that fake a photon due to track misreconstruction. In the following we discuss the various SM processes that give rise to prompt and non-prompt backgrounds ordered according to their severity.

\noindent
\begin{itemize}
\item QCD multijet:

Although the jet faking photon probability is small in the high $p_T$ region of interest of this analysis($\sim 10^{-5}$ as  estimated from our Monte-Carlo Analysis),
the sheer enormity of the cross-section($\sim  millibarns$ above our $p_T$ thresholds) makes this the largest background to the di-photon final state.

  To estimate this background as accurately as possible, we have first generated an EM-enriched sample, which essentially means jets that contain photon-like(EM) objects within themselves. The most common source of jet faking a photon is through $\pi^0$ inside the jet, which decays into two photons. Other meson decays, electron faking photon and fragmentation photons contribute a lesser but non-negligible amount.

We have considered all QCD multijet final states which contain any one of the following objects: photon, electron, $\pi^0$ or $\eta$ mesons (namely the EM-objects). Then we have categorized only those objects which have $p_T > 5$ GeV and are within the rapidity-range $|\eta| < 2.7$, as `seeds'. Then energies and $p_T$ of all the EM-objects within $\Delta R < 0.09 $ around the seed are added with the energy and $p_T$ of the seed. Thus, out of all those EM-objects within a jet, photon candidates are created. If in a QCD multijet event, there are at least two photon candidates with $p_T \gsim$ 30 GeV, those events can in principle fake as a photon with high probability. However, one should also demand a strong isolation around those photon candidates following the isolation criteria described earlier to differentiate between these jet faking photons and actual isolated hard photons.

\item $\gamma +$ jets: This background already has an isolated photon candidate. However, here too, the jets in the final state can fake as photon with a rather small probability($\sim 0.003$ as  estimated from our Monte-Carlo Analysis). But again the large cross-section($\approx 10^5$ pb) of this process makes suppression of the background challenging. For correct estimation of this background, we adopt the same method that has been applied for the multijet background discussed earlier.

\item di-photon: As mentioned above, this background includes production of two photons in the final state through gluon-initiated box diagram, and also via quark-initiated Born diagrams. Although this background gives rise to two isolated hard photons, it does not contribute much due to relatively low cross-section. Demanding a hard $\slashed{E_T}$ and using the fact that the invariant mass of the di-photon pair should peak around the Higgs mass, one can get rid of this background.

\item $Z(\rightarrow \nu \bar{\nu}) h(\rightarrow \gamma \gamma)$: This is an irreducible background for our signal process. This gives rise to sizeable $\slashed{E_T}$, with the invariant mass of the di-photon pair peaking around $m_h$. However, this process has small enough cross-section compared to other backgrounds and proves to be inconsequential in the context of signal significance. 
\end{itemize}

A few comments are in order before we delve deeper into our analysis. Some studies in the recent past have considered, Higgs production through higher dimensional operators, based on the di-photon signal~\cite{Carpenter:2013xra}. However, the role of backgrounds from QCD multijets has not been fully studied there. Our analysis in this respect is more complete. Also, $p p \rightarrow W^+ W^-$, too, in principle lead to substantial $\slashed{E_T}$ + two ECAL hits with electrons from both the $W$s being missed in the tracker. We can neglect, such fakes because (a) the event rate is double-suppressed by electronic branching ratios, (b) demand on the invariant mass helps to reduce the number of events and (c) two simultaneous fakes by energetic electrons is relatively improbable.

\subsection{$b \bar b +\slashed{E_T}$ channel}

{\bf Signal:}  The $b \bar b + \slashed{E_T}$ channel resulting in hadronic final states, poses a seemingly tougher challenge, as compared to the di-photon final state. However, the substantial rate in this channel creates an opportunity to probe the mono-Higgs+MET signal, if backgrounds can be effectively handled. Searches in this channel have been carried out by both CMS~\cite{CMS-PAS-EXO-16-050,Sirunyan:2017hnk,Sirunyan:2018gdw} and ATLAS~\cite{ATLAS-CONF-2016-019,Aaboud:2017yqz,ATLAS:2018bvd} experiments.

One demands two energetic $b$-tagged jets with $p_T > 20$ GeV as trigger, along with considerable $\slashed{E_T}$. The efficiency of tagging $b$-jets can be optimized to approximately 70\% if the $p_T$-range is further tweaked, something that we need to do for background suppression as well.

It is clear from Figure~\ref{crosssec} that here too the resonance region ($m_\chi \gsim \frac{m_h}{2}$) offers the best signal prospect, as in the $\gamma \gamma$ case. 
The large cross-section in the resonant region enables one to explore $\lambda_{\phi \chi} \lsim 6$ in a cut-based analysis, which can be further improved in a machine learning approach, as we shall see below. This underscores the importance of the $b \bar b$ channel, vis-a-vis di-photon. For such $\lambda_{\phi\chi}$, one can probe upto $\Lambda \approx 8$ TeV.
On the other hand it is also possible to probe heavier DM states in this channel. Although higher $m_{\chi}$ will imply lower yield (see Figure~\ref{crosssec}), the kinematic distributions such as that in $\slashed{E_T}$ will have more discriminating power compared to the resonant region. Therefore, the $b \bar b$ channel also helps us probe larger mass windows for the DM particle, compared to the di-photon case. Keeping these points in mind we chose a few benchmark points listed in Table~\ref{bpbb}, satisfying all the aforementioned constraints. Here too, the choice of benchmarks are governed by the prospect of detectability at the collider. Here we choose BP5 to be the best possible case where dark matter mass is in the resonant region and the quartic coupling $\lambda_{\Phi \chi}$ is close to the perturbativity limit yielding the largest possible production cross-section. One should note that BP5 is exactly same as BP2 of $\gamma \gamma + \slashed{E_T}$ analysis discussed earlier. BP4 focusses on the prospect of the $b \bar b + \slashed{E_T}$ channel in terms of dark matter mass. Therefore a heavier DM has been chosen, keeping the quartic coupling same as BP5. BP6 probes lower quartic couplings compared to BP4 and BP5 when the dark matter mass continues to remain in the resonant region.


\begin{table}[!hptb]
\begin{center}
\begin{footnotesize}
\begin{tabular}{| c | c | c | c |}
\hline
Benchmarks & $m_{\chi}$ & $\lambda_{\Phi \chi}$ & $\Lambda$ \\
\hline
BP4 & 120.0 & 12.0  & 5 TeV \\
\hline
BP5 & 64.0 & 12.0 & 5 TeV \\
\hline
BP6 & 64.0 & 6.0 & 8 TeV \\
\hline
\end{tabular}
\end{footnotesize}
\caption{The benchmark points for $b \bar b + \slashed{E_T}$ final state.}
\label{bpbb}
\end{center}
\end{table}

Having chosen benchmark points for signal we proceed to analyze the corresponding backgrounds.

{\bf Background:} We list the dominant backgrounds for this channel in the following.

\begin{itemize}
\item $t \bar t$ + single top: The major background for the $ b \bar b$ channel comes from $t \bar t$ production at the LHC. The hadronic, semileptonic and leptonic decays of $t \bar t$ produce $b \bar b$ pairs in the final state. The contribution is largest for the semileptonic decay of $t \bar t$. It has substantial production rate and is also source of substantial $\slashed{E_T}$. A minor contribution comes from the purely leptonic $t \bar t$ decay as well. It also has $\slashed{E_T}$ in the final state from two neutrinos coming from leptonic $W$ decay. However, a veto on $p_T$ of leptons $> 10$ GeV reduces this background at the selection level itself, whereas the semileptonic $t \bar t$ background is less affected by such veto. The hadronic $t \bar t$ background has the largest cross-section among all $t \bar t$ backgrounds, but in this case the source of $\slashed{E_T}$ is mismeasurement of jet energy. A full simulation shows that the non-leptonic $t \bar t$ decay background plays a sub-dominant role. The single top background is also taken into account, but its contribution is rather small compared to the semileptonic and leptonic $t \bar t$ because of its much smaller cross-section. 

\item $V+$ jets: The next largest contribution to the background, in our signal region comes from $V+$ jets ($V = W, Z)$ production. These processes have large cross-sections ($\approx 10^4$ pb) and also have significant sources of $\slashed{E_T}$ through the semileptonic decays of the weak gauge bosons. However, this background depends on the simultaneous mistagging of two light jets as $b$-jets. The double-mistag probability is rather small for these backgrounds ($\approx 0.04\%$ as estimated from our Monte-Carlo simulation). It is worth mentioning here, that the contribution of $W+$jets is found to be sub-dominant compared to $Z+$jets. The main reason behind this is the presence of larger $\slashed{E_T}$ in the latter case and also the suppression of the former by the lepton veto.

\item QCD $b \bar b$: One major drawback of the $b \bar b$ channel is the presence of QCD $b \bar b$ production of events which has large cross-section ($\approx 10^5$ pb). The nuisance value of this background, however,depends largely on $\slashed{E_T}$ coming from jet-energy mismeasurement.  
On applying a suitable strategy which we will discuss in the next section (see Table 5), we find that this background becomes sub-dominant to those from $t \bar t$ and $Z+$jets processes. 

\item $Z(\rightarrow \nu \bar \nu)h(\rightarrow b \bar b)$: Similar to the $\gamma \gamma$ case, this background too is irreducible. The $\slashed{E_T}$ and invariant mass of the $b \bar b$ system are also similar to the signal processes. However the smallness of its cross-section ($\approx 100$ fb) makes this background less significant compared to all other backgrounds described earlier.
\end{itemize}

\section{Collider Analysis: Cut-based }  
\label{sec5}
\subsection{$\gamma \gamma +\slashed{E_T}$ channel}

The discussion in the foregoing section convinces us that it is worthwhile to look at the $\gamma \gamma + \slashed{E_T}$ channel because of the `clean' di-photon final state.
A refinement of the already adopted strategies are thus expected to be useful in the high-luminosity LHC.    
Our rectangular cut-based analysis for this final state is in this spirit. 

Events for the signals and most of the corresponding backgrounds have been generated using Madgraph@MCNLO~\cite{Alwall:2014hca} and their cross-sections have been calculated at the next-to-leading order(NLO). We have used the nn23lo1 parton distribution function. The di-jet and $\gamma +$ jet backgrounds are generated directly using PYTHIA8~\cite{Sjostrand:2006za}. PYTHIA8 has been used for the showering and hadronization and the detector simulation has been taken care of by Delphes-3.4.1~\cite{deFavereau:2013fsa}. Jets are formed by the built-in Fastjet~\cite{Cacciari:2006sm} of Delphes.

\begin{figure}[!hptb]
\centering
\includegraphics[width=7.5cm, height=5.5cm]{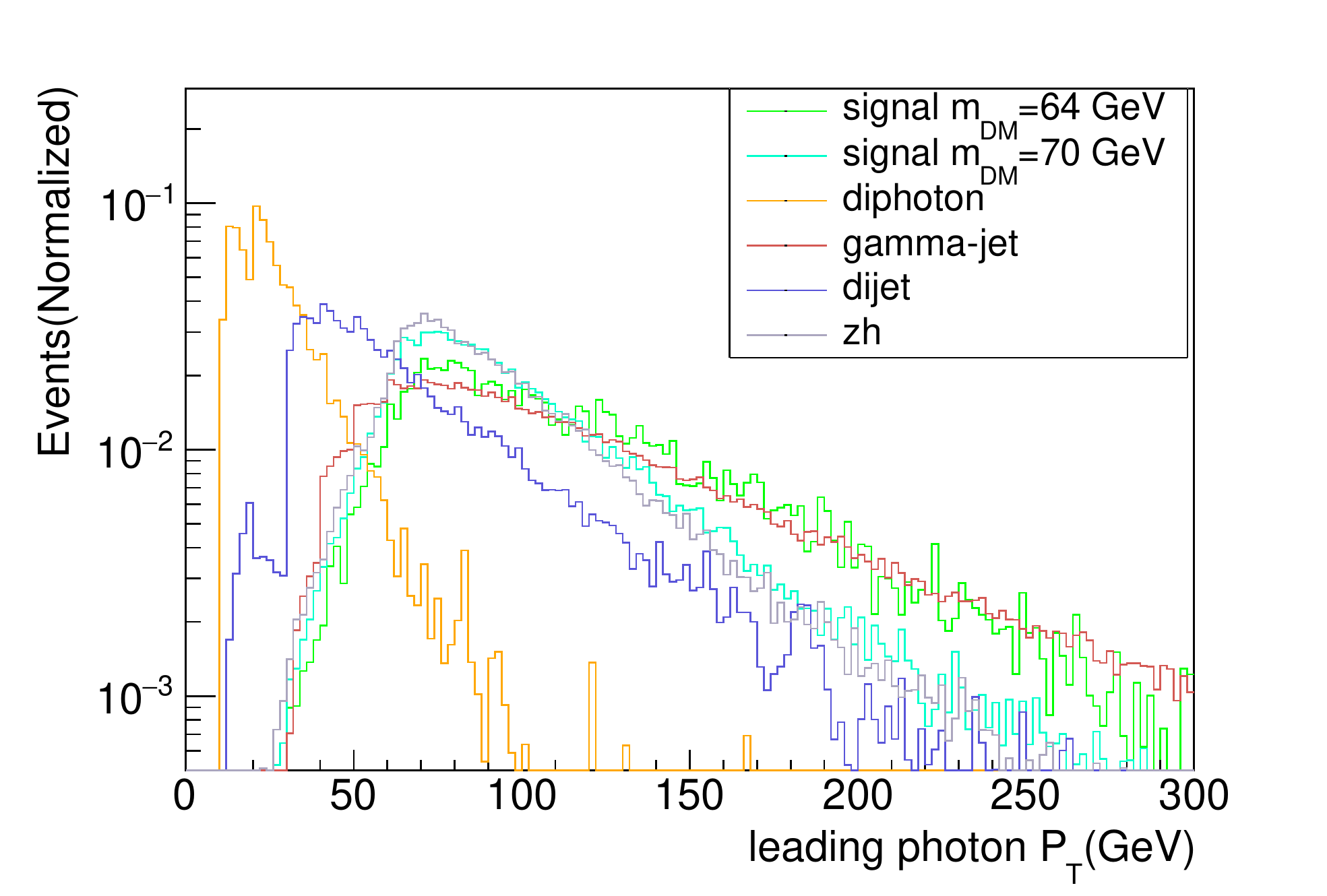} 
\includegraphics[width=7.5cm, height=5.5cm]{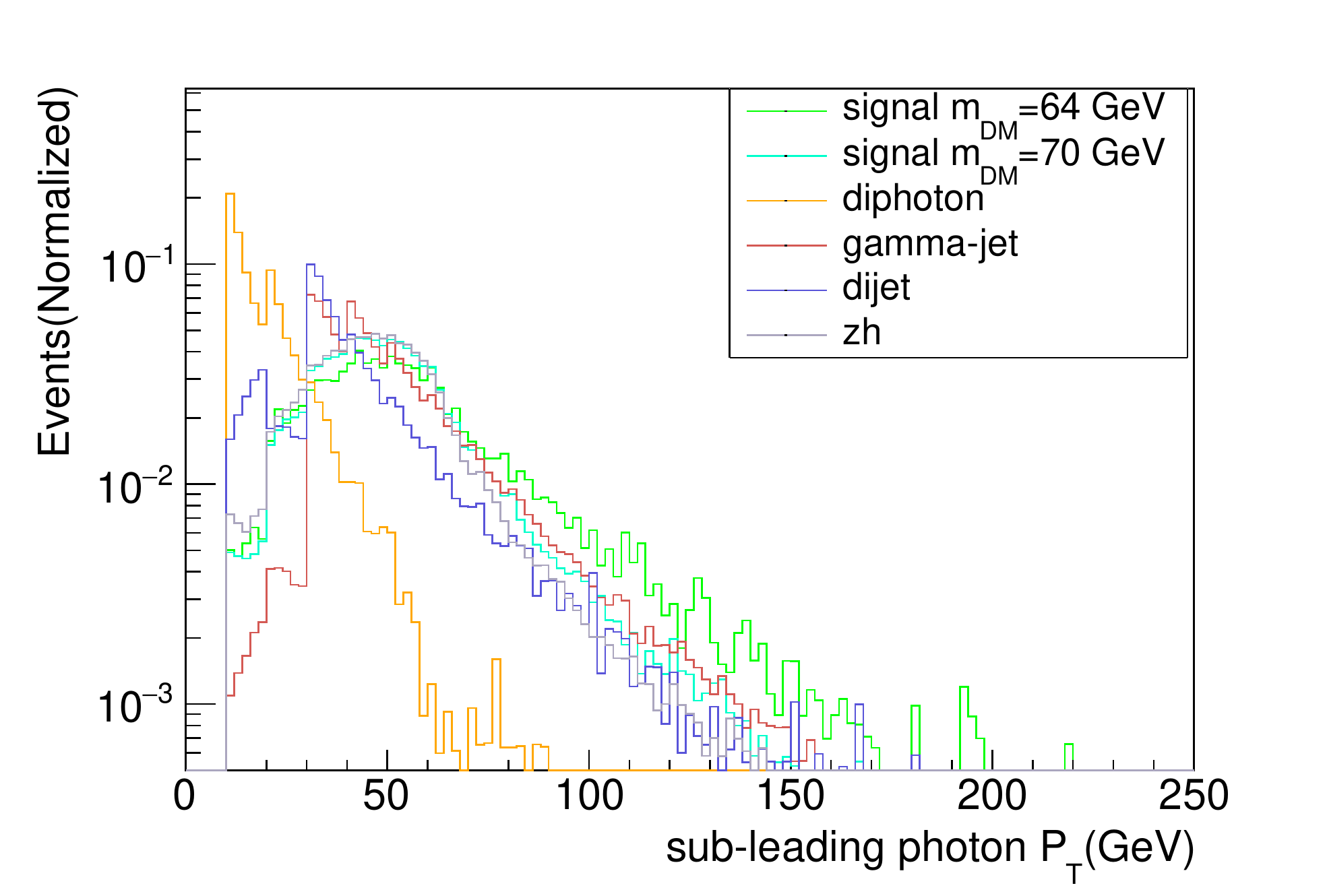}
\caption{Distribution of transverse momenta of the leading~(left) and sub-leading~(right) photons. }
\label{ptgamma}
\end{figure}

We will discuss the results of our cut-based analysis for a few benchmarks presented in Table~\ref{bpgam} which are allowed by all the constraints mentioned earlier. We will first identify variables which give us desired separation between the signal and backgrounds. We present in Figure~\ref{ptgamma} the distribution of the transverse momenta of the leading and sub-leading photon for the signal and all the background processes. The signal photons are recoiling against the dark matter and therefore are boosted. On the other hand, in case of di-jet events the photons are part of a jet, and it is unlikely that those photons will carry significant energy and $p_T$ themselves. In case of $\gamma$ + jet background at least one photon is isolated and it can carry considerable $p_T$. Therefore the $p_T$ distribution of the leading photon is comparable with the signal in this case, while for the sub-leading photon, which is expected to come from a jet, it falls off faster as expected.

\begin{figure}[!hptb]
\centering
\includegraphics[width=7.5cm, height=5.5cm]{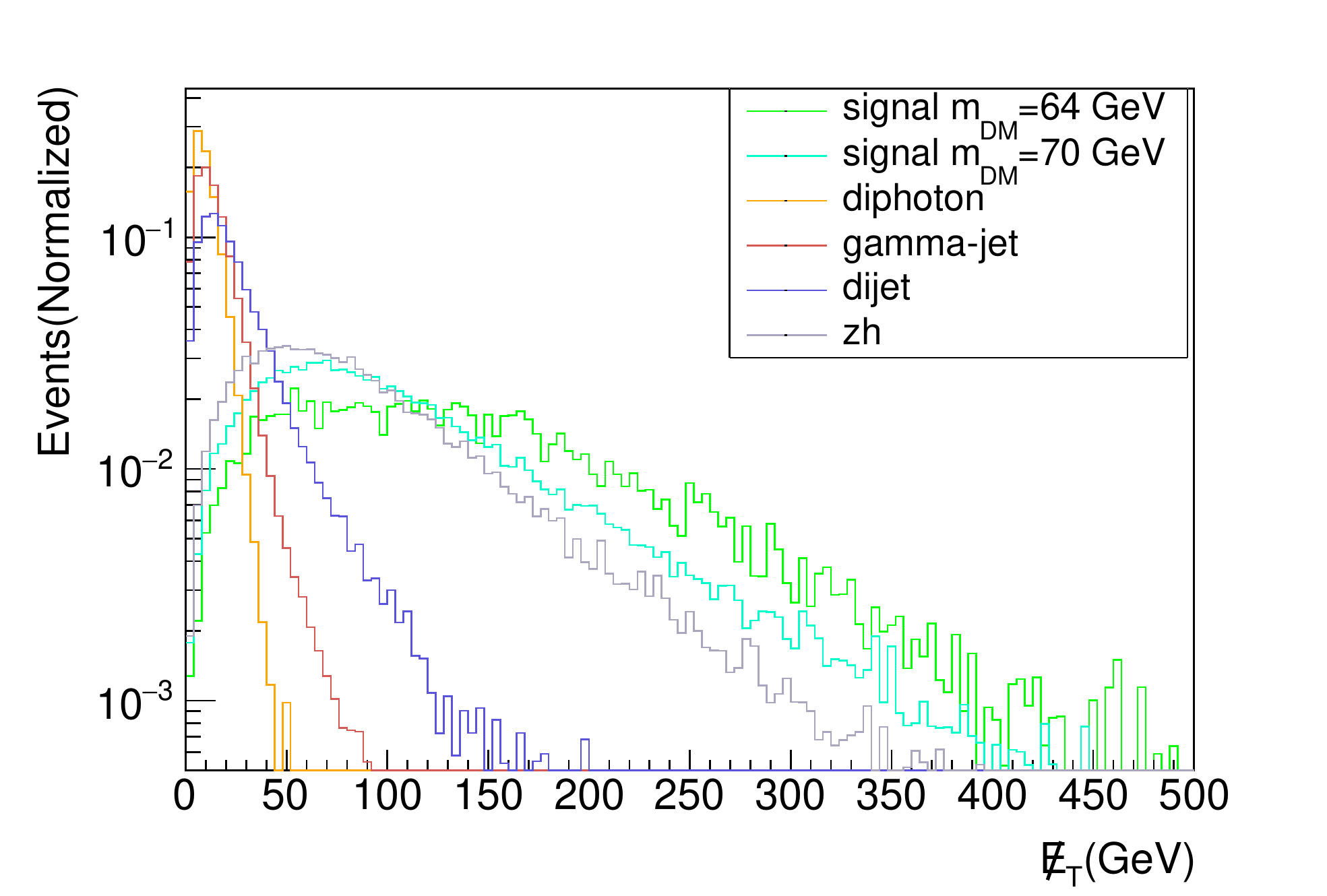} 
\includegraphics[width=7.5cm, height=5.5cm]{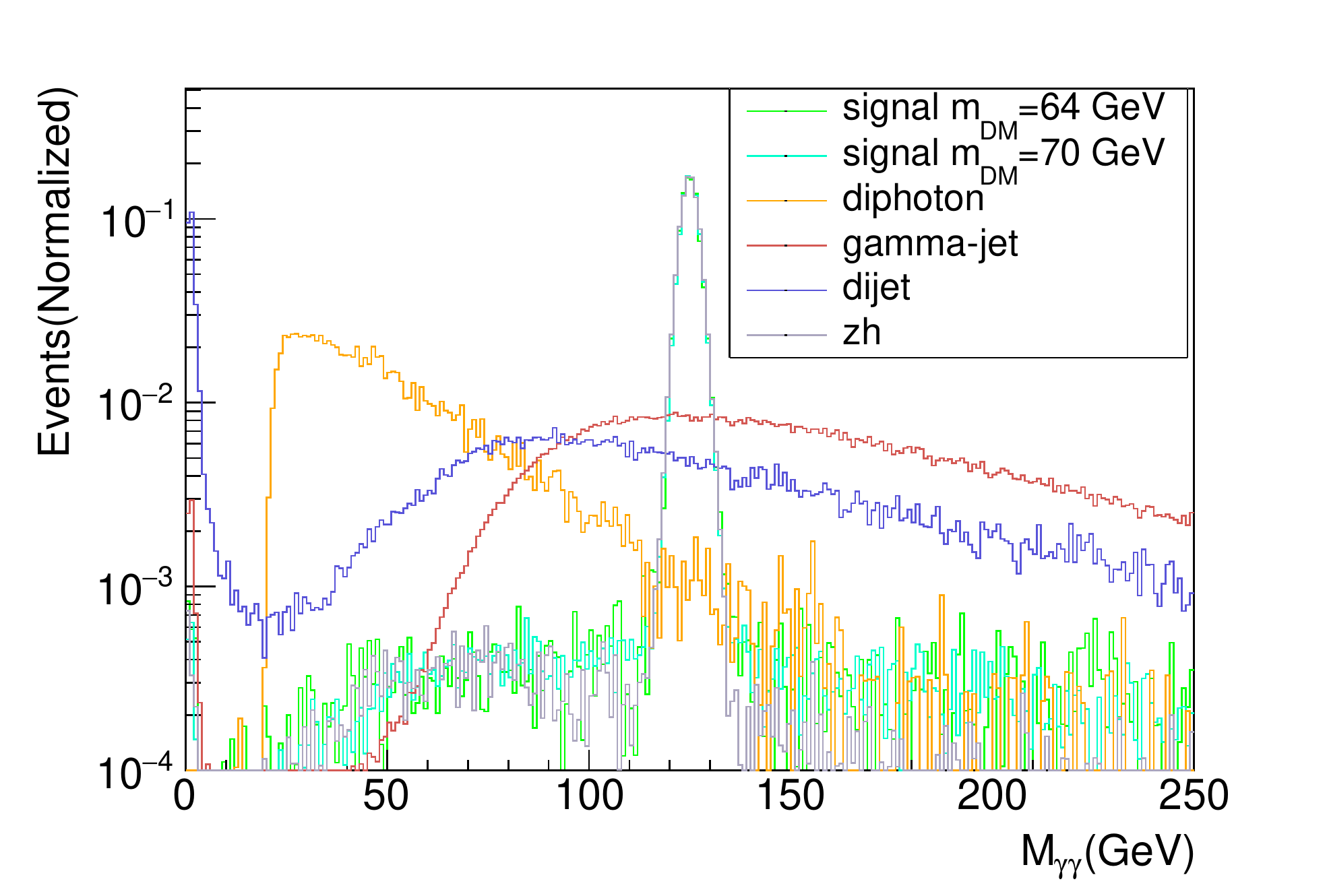}
\caption{$\slashed{E_T}$ distribution~(left) and invariant mass distribution of leading and sub-leading~(right) photons. }
\label{metinv}
\end{figure}

In Figure~\ref{metinv}~(left), we plot the $\slashed{E_T}$ distribution for signal and background processes. $\slashed{E_T}$ distributions for di-jet, $\gamma$+jet and di-photon show that these background events will have much less $\slashed{E_T}$ compared to the signal and $Zh$ background. In the background events involving jets the source of missing energy is via the mismeasurement of jet energy or from the decay of some hadron inside the jet. $\gamma \gamma$ background naturally shows lowest $\slashed{E_T}$ contribution owing to the absence of real $\slashed{E_T}$ or jets in the final state. On the other hand, the signal and $Zh$ background have real source of $\slashed{E_T}$. It is evident that $\slashed{E_T}$ observable will play a significant role in signal-background discrimination. In Figure~\ref{metinv}~(right), we have shown the invariant mass of the di-photon pair. In case of signal and $Zh$ background the $m_{\gamma\gamma}$ distribution peaks at Higgs mass and for all other backgrounds no such peak is observed. This distribution is also extremely important for background rejection.

\begin{figure}[!hptb]
\centering
\includegraphics[width=7.5cm, height=5.5cm]{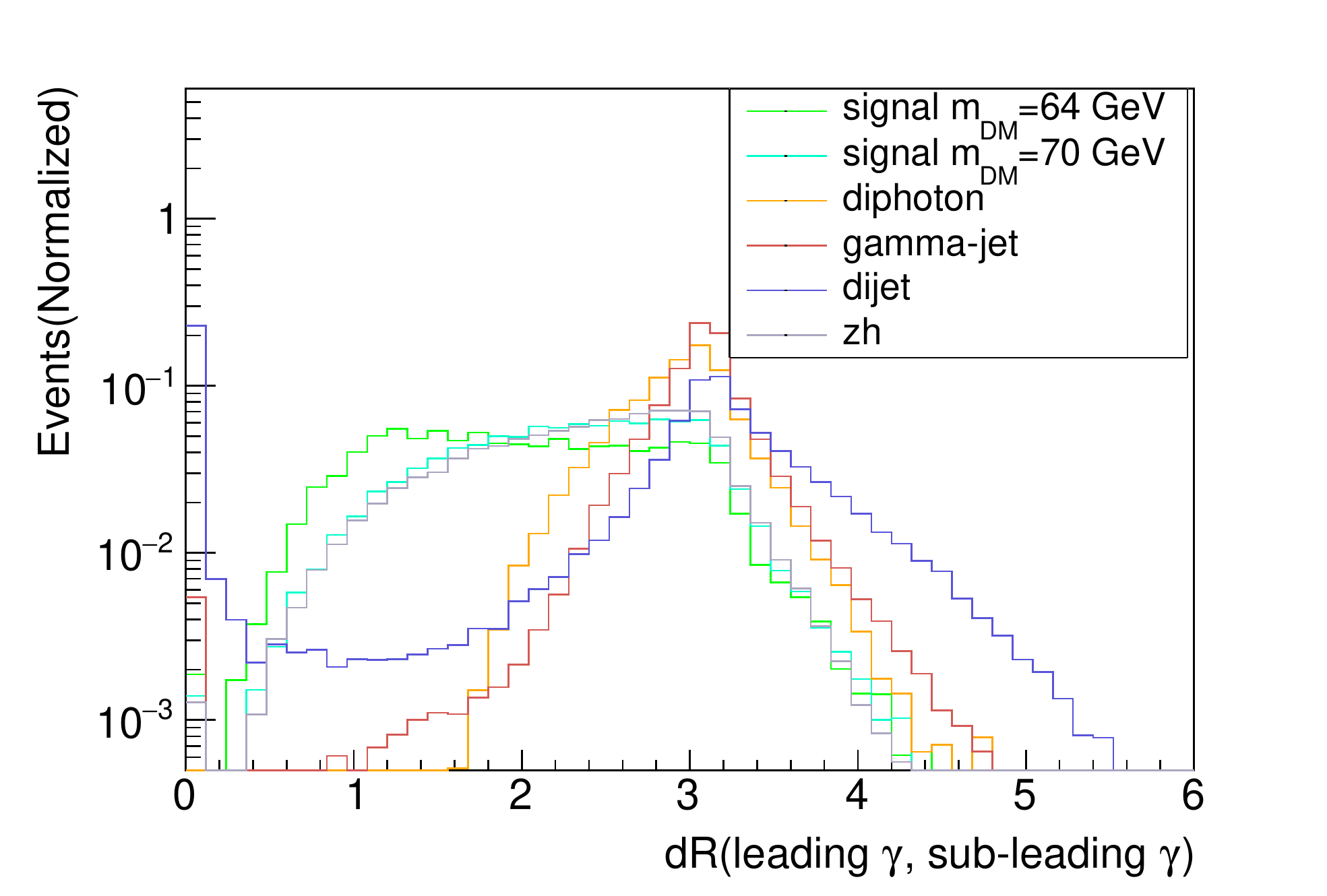} 
\includegraphics[width=7.5cm, height=5.5cm]{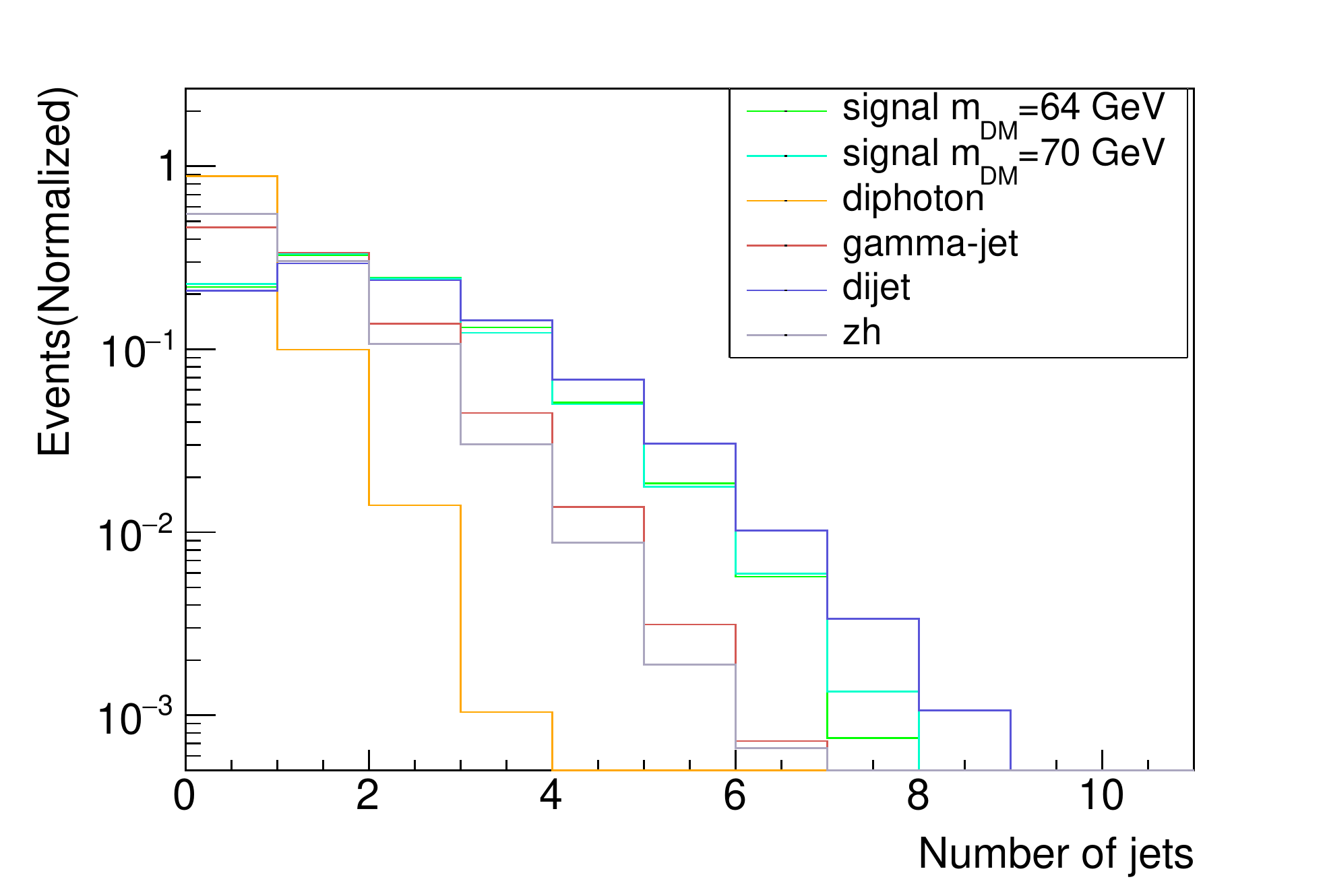}
\caption{$\Delta R$ distribution between the leading and sub-leading photons (left) and distribution of number of jets~(right). }
\label{dphinjet}
\end{figure}

\begin{figure}[!hptb]
\centering
\includegraphics[width=7.5cm, height=5.5cm]{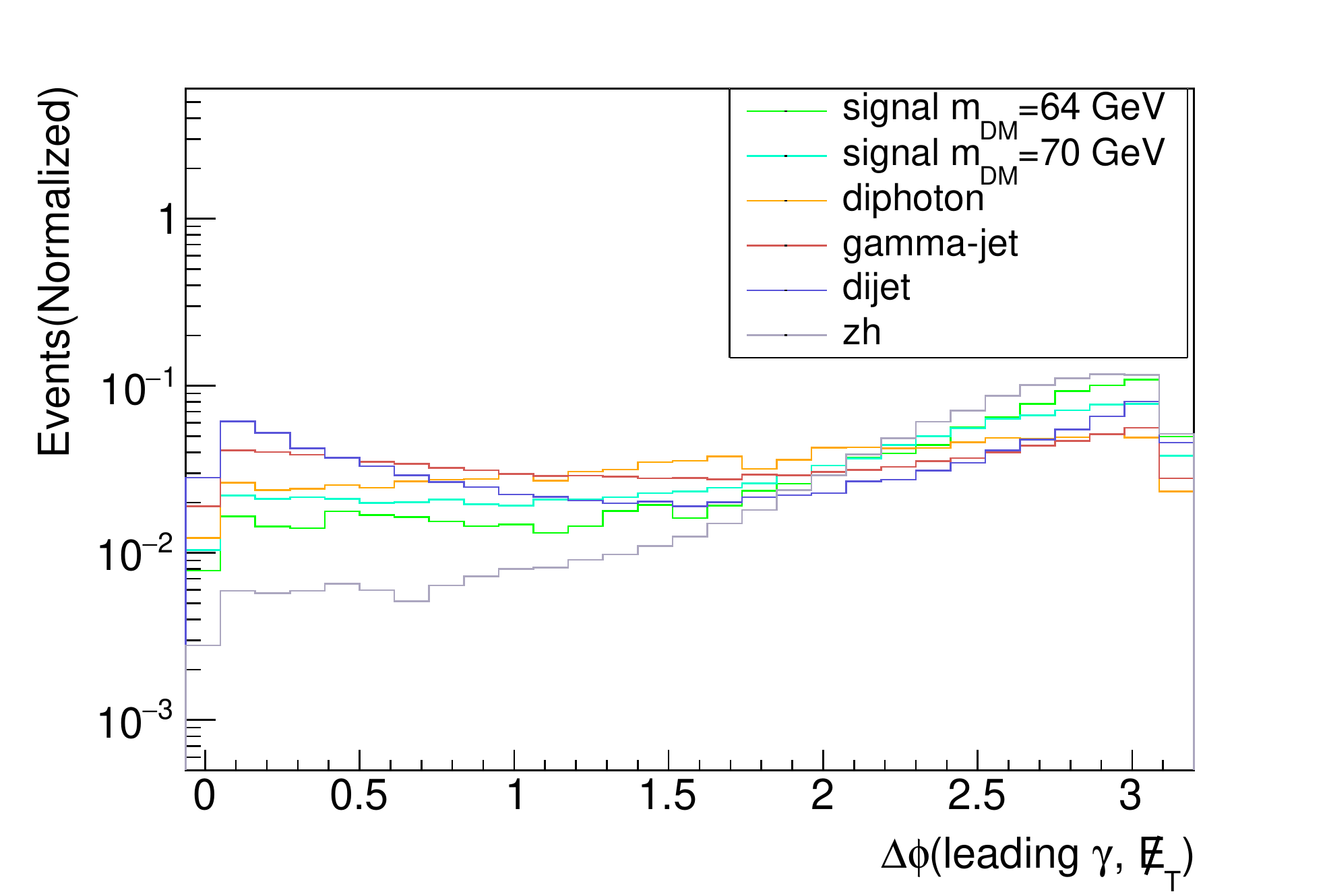} 
\includegraphics[width=7.5cm, height=5.5cm]{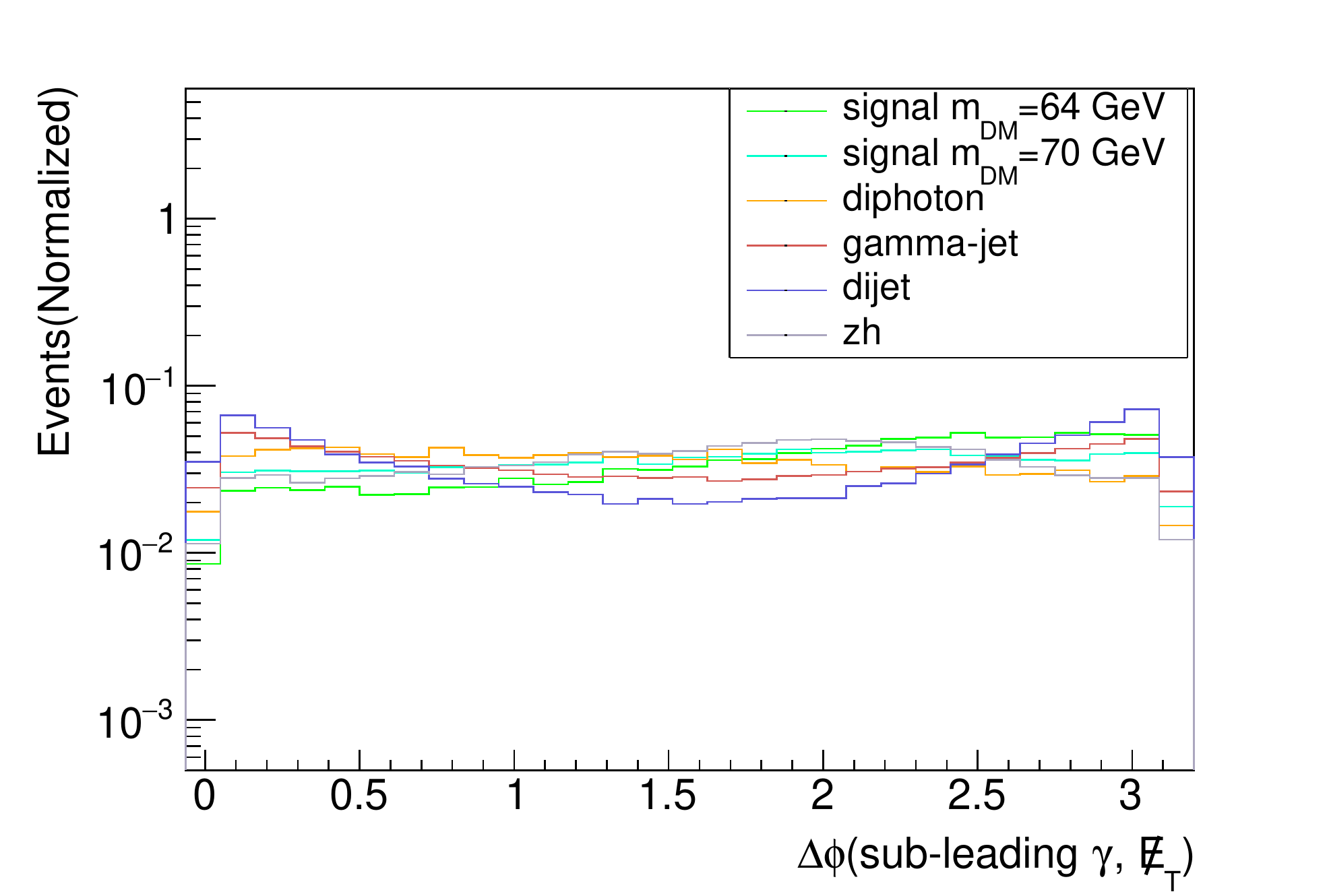}
\caption{$\Delta \phi$ distribution between the $\slashed{E_T}$ and leading~(left) and sub-leading~(right) photons. }
\label{dphiphomet}
\end{figure}

In Figure~\ref{dphinjet}~(left) we plot the distribution of $\Delta R$ between the two photons for signal and background. We can see from the figure that in case of di-jet and $\gamma + $jet background there is a peak at $\Delta R \approx 0$, resulting from the cases where the two photons have come from a single jet. However the di-jet and  $\gamma + $jet events have a second peak too because of the events where each photon is part of a single jet and therefore the two photons are back to back. However, we do not really observe any peak in the $\Delta R$ distribution of the signal or $Zh$ background because in these cases the di-photon system is exactly opposite to the dark matter pair. Next we plot the jet multiplicity distribution for signal and background in Figure~\ref{dphinjet}~(right). 

In Figure~\ref{dphiphomet} (left) and (right), we have plotted the $\Delta \phi$ distribution between the $\slashed{E_T}$ and the leading and sub-leading photon respectively. We know that in case of signal and the $Zh$ background the di-photon system is exactly opposite to the dark matter system. Therefore the photons from the Higgs boson decay tend to be mostly opposite to the $\slashed{E_T}$. On the other hand in case of di-jet or $\gamma+$jet events, the the $\slashed{E_T}$ is aligned with either of the jets for aforementioned reasons. This behaviour is visible from the Figure~\ref{dphiphomet}. 

We would like to remind the reader that no special strategy has been devised for the irreducible $Zh$ background. This is because of its low rates compared to both the signal and the other background channels.

\bigskip

\paragraph{Results:}

\bigskip

Having discussed the distributions of the relevant kinematical variables we go ahead to analyze the signal and background events after applying appropriate cuts over the observables. Over and above the basic event selection criteria, the following cuts were applied. 

\begin{itemize}

\item Cut 1: $p_T$ of the leading(sub-leading) photon $> 50(30)$ GeV 
\item Cut 2: $\Delta R$ between two photons $> 0.3$
\item Cut 3: $\Delta \phi$ between leading(sub-leading) photon and $\slashed{E_T} > 0.3(0.3)$
\item Cut 4: $\slashed{E_T} > 100$ GeV
\item Cut 5: $\Delta \phi$ between leading(sub-leading) jet and $\slashed{E_T} > 0.4(0.4)$ 
\item Cut 6: $115$ GeV $< m_{\gamma\gamma} < 135$ GeV
\end{itemize} 

A clarification is in order on the way in which we apply the isolation requirement on each photon.  We have imposed the requirement that the total scalar sum of transverse momenta of all the charged and neutral particles within $\Delta R < 0.5$ of the candidate photon can not be greater than 12\% of the $p_T$ of the candidate photon i.e
$\frac{\sum_{i} p_T^i(\Delta R < 0.5)}{p_T^{\gamma}} < 0.12$. Thereafter we have plotted the isolation probability of a photon, defined by this criterion, as a function of $p_T$.
We have then multiplied the $p_T$-weighted isolation probability with all the events surviving after applying the previously mentioned cuts. We call this criterion Cut7 which goes somewhat beyond the rectangular cut-based strategy.

\begin{table}[!hptb]
\scriptsize{
\begin{tabular}{| c | c | c | c | c | c | c | c | c |}
\hline
Datasets & Xsec(pb) & Cut1 & Cut2 & Cut3 & Cut4 & Cut5 & Cut6 & Cut7   \\
\hline
BP1 & $3.0\times 10^{-4}$ & 56.2\% & 56.0\% & 48.0\% & 23.8\% & 21.6\% & 19.3\% & 14.7\%   \\
\hline
BP2 & $4.4\times 10^{-4}$ & 57.2\% & 57.0\% & 51.5\% & 33.3\% & 30.1\% & 27.3\% & 21.0\%   \\
\hline
BP3 & $1.9\times 10^{-4}$ & 57.2\% & 57.0\% & 51.5\% & 33.3\% & 30.1\% & 27.3\% & 21.0\%   \\
\hline
$jj$ & $7.8\times 10^{8}$ & $8.9\times 10^{-4}$\% & $7.0\times 10^{-4}$\% & $4.6\times 10^{-4}$\% & $3.6\times 10^{-6}$\% & $1.1\times 10^{-6}$\% & $1.5\times 10^{-7}$\% & $1.2\times 10^{-10}$\%   \\
\hline
$\gamma j$ & $1.5\times 10^5$ & 0.09\% & 0.09\% & 0.07\% & $4.2\times 10^{-5}$\% & $2.5\times 10^{-5}$\% & $1.6\times 10^{-6}$\% & $4.8\times 10^{-8}$\%   \\
\hline
$\gamma \gamma$ & 410 & 53.0\% & 53.0\% & 44.1\% & $2.7\times 10^{-4}$\% & $2.7\times 10^{-4}$\% & $6.7\times 10^{-5}$\% & $6.0\times 10^{-5}$\%   \\
\hline
$Zh$ & $2.7\times 10^{-4}$ & 44.2\% & 44.2\% & 40.5\% & 14.5\% & 13.6\% & 13.1\% & 11.6\%   \\
\hline
\end{tabular}
\caption{Signal and background efficiencies in the $\gamma \gamma + \slashed{E_T}$ final state after applying various cuts at 13 TeV. The cross-sections are calculated at NLO.}
\label{tablecutflow}
}
\end{table}


Table~\ref{tablecutflow}, indicates the cut-efficiencies of various kinematic observables. We can see from this table that $\slashed{E_T}$, $m_{\gamma\gamma}$ and the isolation criterion turn out to be most important in the separation of the signal from the background. Having optimized cut values, we calculate the projected significance (${\cal S})$ for each benchmark point for the 13 TeV LHC with 3000 $fb^{-1}$ in Table~\ref{significancegamgam}. The significance ${\cal S}$ is defined as 

\begin{equation}
{\cal S} = \sqrt{2 [(S+B) \text{Log}(1+\frac{S}{B}) - S]}
\label{significance}
\end{equation}

Where $S$ and $B$ are the number of signal and background events surviving the 
succession of cuts.

\begin{table}[!hptb]
\begin{center}
\begin{tabular}{| c | c |}
\hline
BP & $
{\cal S}$   \\
\hline
BP1  & 2.4 $\sigma$  \\
\hline
BP2  & 4.8 $\sigma$  \\
\hline
BP3  & 2.1 $\sigma$  \\
\hline
\end{tabular}
\caption{Signal significance for the benchmark points at 13 TeV with ${\cal L}$ = 3000 $fb^{-1}$ in the $\gamma \gamma + \slashed{E_T}$ channel, after using the cuts listed in Table~\ref{tablecutflow}. }
\label{significancegamgam}
\end{center}
\end{table}

From Table~\ref{significancegamgam}, one can see that even with cut-based analysis it is possible to achieve $\approx 5 \sigma$ significance for BP2 which has the largest production cross-section. The other benchmarks with heavier dark matter mass or small quartic coupling do not perform very well. One should note that although we have identified the strongest classifier observables through our cut-based analysis and used them, it is possible to make use of the weaker classifiers too if we go beyond cut-based set-up, which will precisely be our goal with machine learning in the next section. We hope to achieve some improvement over the results quoted in Table~\ref{significancegamgam}.

\subsection{$b \bar b +\slashed{E_T}$ channel}


We present here the analysis for the $b \bar b +\slashed{E_T}$ final state in the context of high luminosity LHC.
The signal and background events (except QCD $b \bar b$) are generated using Madgraph@MCNLO~\cite{Alwall:2014hca} and showered through PYTHIA8~\cite{Sjostrand:2006za}. The QCD $b \bar b$ background is generated directly using PYTHIA8. The detector simulation is performed by Delphes-3.4.1~\cite{deFavereau:2013fsa}, the jet formation is taken care of by the built-in Fastjet~\cite{Cacciari:2006sm} of Delphes.

\begin{figure}[!hptb]
\centering
\includegraphics[width=7.5cm, height=5.5cm]{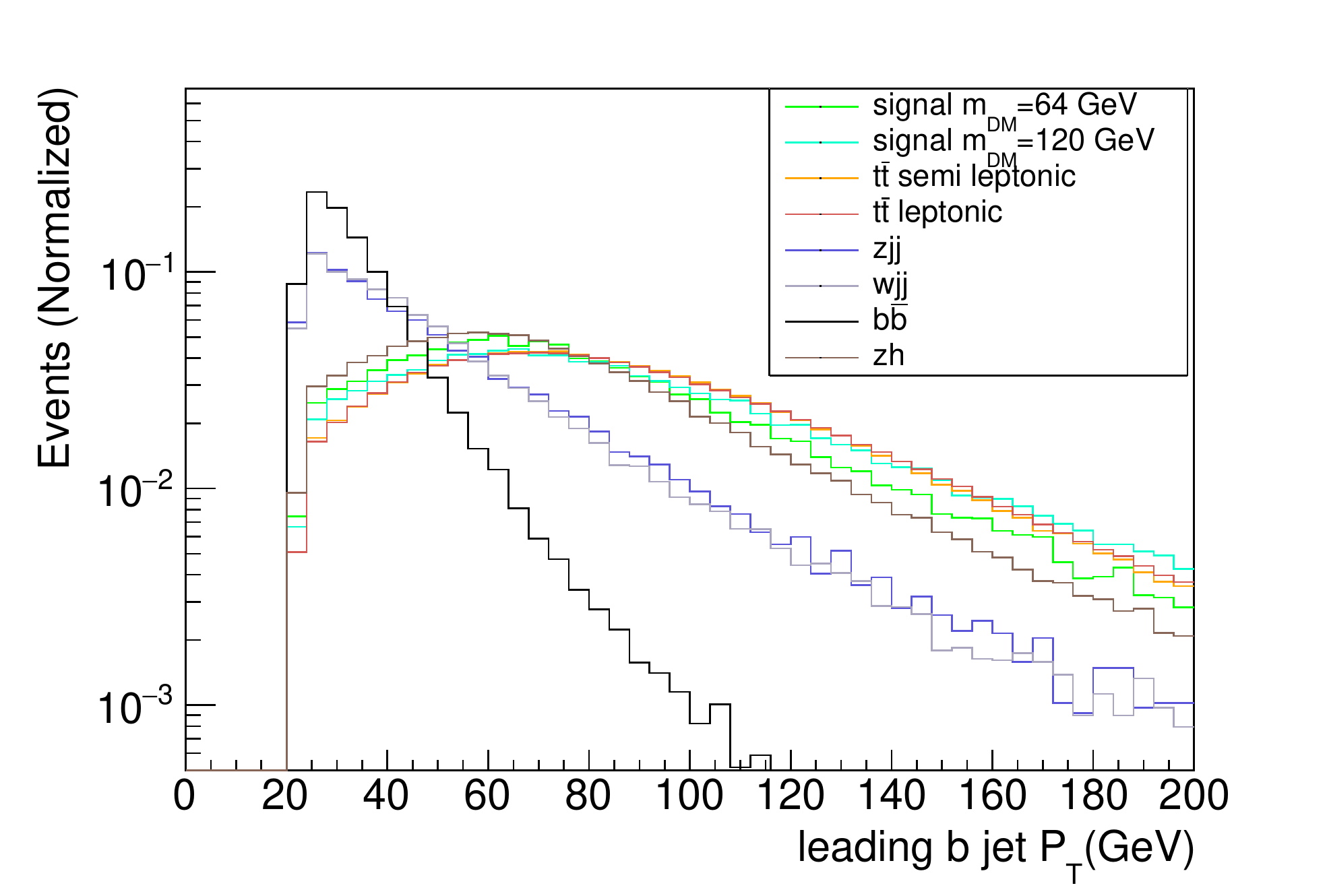} 
\includegraphics[width=7.5cm, height=5.5cm]{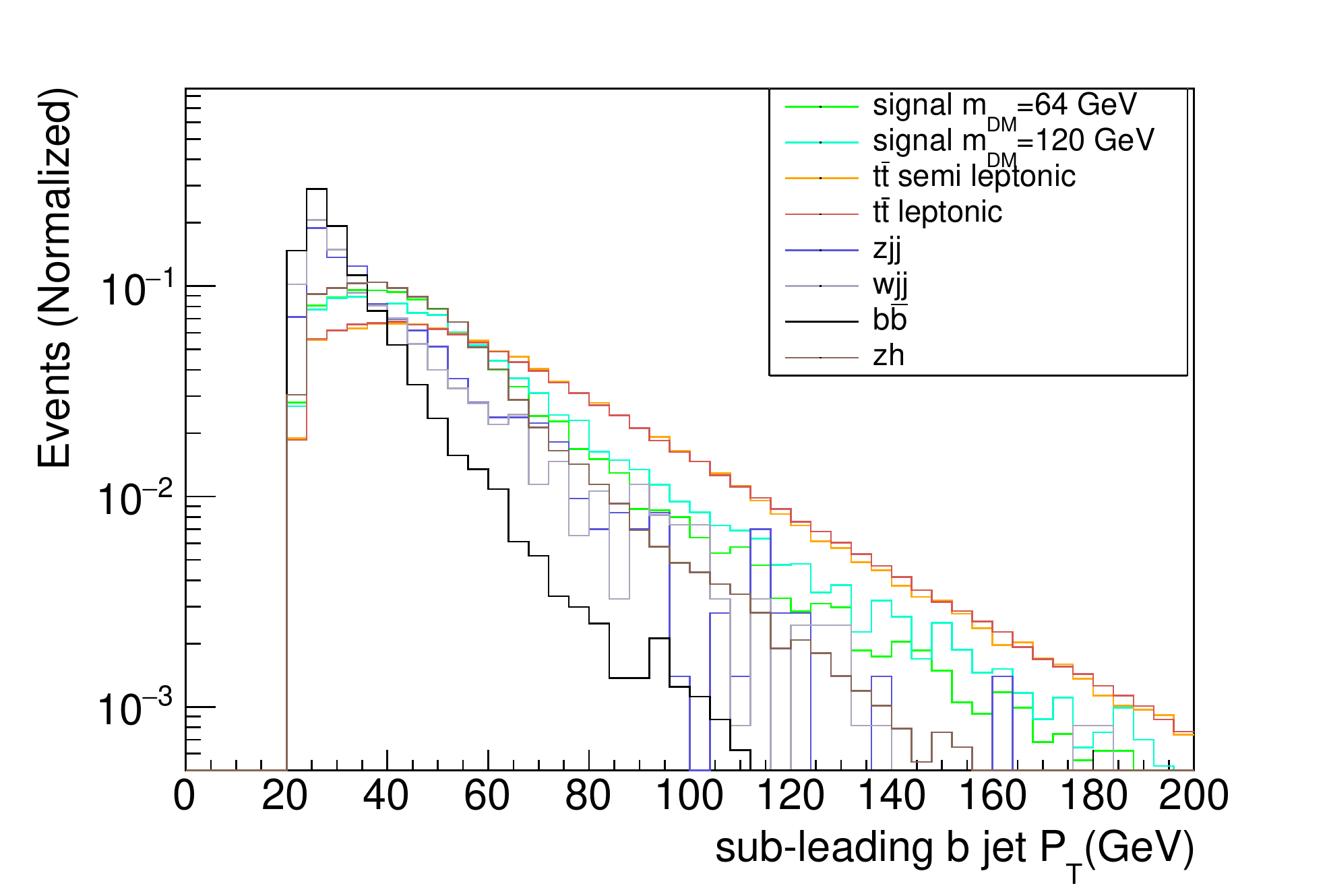}
\caption{Distribution of the transverse momenta of the leading~(left) and sub-leading~(right) $b$-jets. }
\label{ptb1b2}
\end{figure}

We proceed to discuss the kinematic variables which yield significant signal-background separation. In Figure~\ref{ptb1b2}, we plot the $p_T$ distribution of the leading and sub-leading $b$-jets for the signal and all the background processes. We can see from the figures that for the signal, the $p_T$ distribution of the $b$ jets peak at a higher value compared to the QCD $b \bar b$ and $V+$jets background. This behaviour is expected since the $b \bar b$ system is recoiling against a massive DM particle in case of signal. The distribution of $p_T$ of leading and sub-leading $b$-jets from $t \bar t$ backgrounds however peak at a similar region as the signal. The $b$-jets in those cases come from the decay of top quarks and are therefore boosted. Guided by the distributions we put appropriate cut on the transverse momenta of the leading and sub-leading $b$-jets in our cut-based analysis.

\begin{figure}[!hptb]
\centering
\includegraphics[width=7.5cm, height=5.5cm]{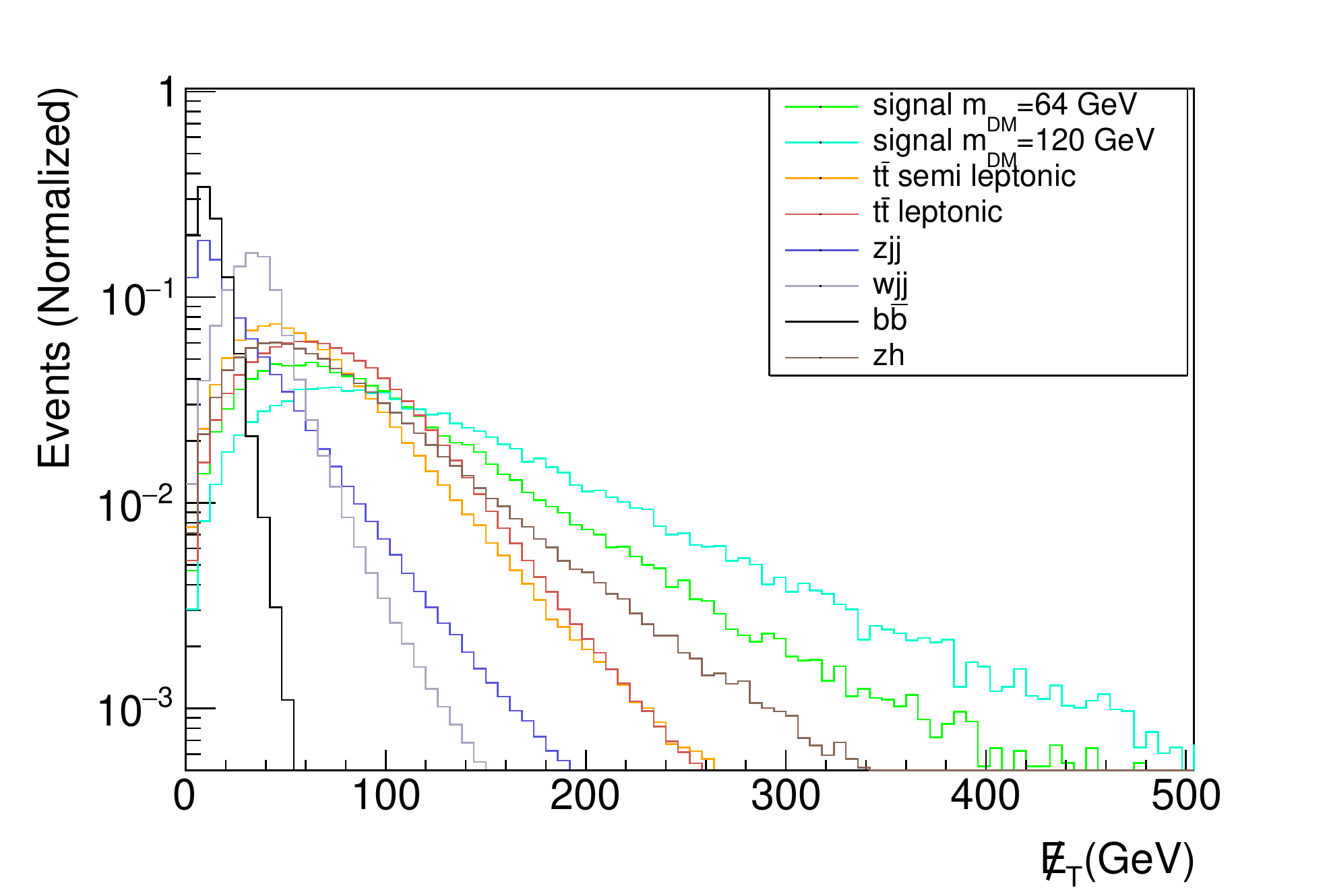} 
\includegraphics[width=7.5cm, height=5.5cm]{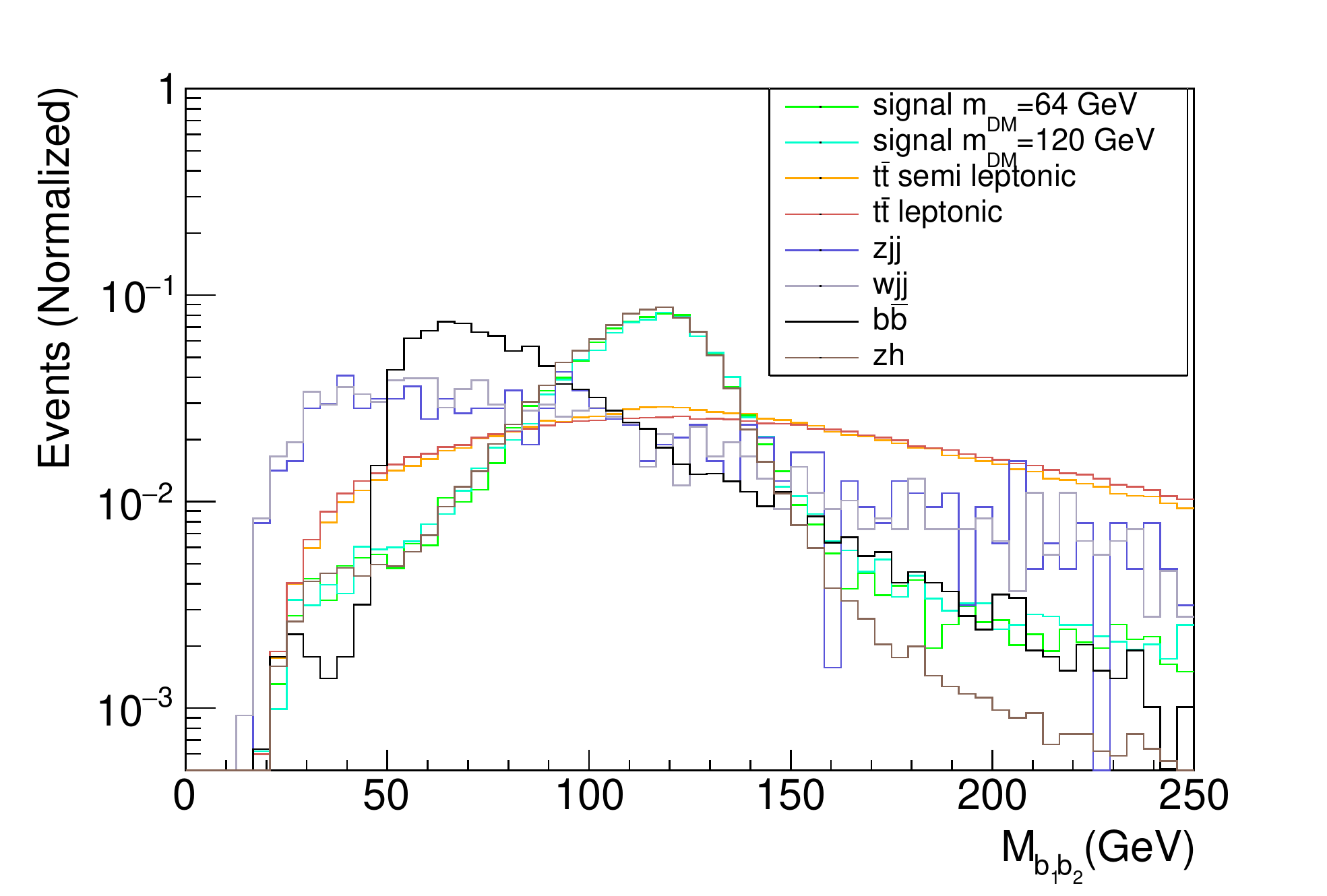}
\caption{$\slashed{E_T}$ distribution~(left) and invariant mass distribution of leading and sub-leading~(right) $b$ jets. }
\label{metinvbb}
\end{figure}

In Figure~\ref{metinvbb} (left) we plot the $\slashed{E_T}$ distribution for signal and background processes. We can see that the QCD background produces the softest $\slashed{E_T}$ spectrum. The reason behind this is that there is no real source of $\slashed{E_T}$ for this final state. It is mainly the mismeasurement of the visible momenta of the jets, which leads to $\slashed{E_T}$.
Though large $\slashed{E_T}$ arising from such mismeasurement contributes mostly to the tail of the distribution, the sheer magnitude of the cross-section can still constitute a menace.
However, a strong $\slashed{E_T}$ cut helps us reduce the $b \bar b$ background to a large extent, as will be clear from the cut-flow analysis in the next subsection. The $\slashed{E_T}$ peaks at much lower values in case of $V$+jets as well. However, $t \bar t$ and $Zh$ background also produce large enough $\slashed{E_T}$, although less than our signal. Therefore a hard $\slashed{E_T}$ cut enhances the signal background separation. In Figure~\ref{metinvbb} (right), we plot the invariant mass distribution of two $b$-jets. In case of signal and $Zh$ background it peaks at Higgs mass, whereas for all other backgrounds it falls off rapidly. It is evident that a suitable cut on the invariant mass of the $b$-jet pair will also be effective in reducing the background.

\begin{figure}[!hptb]
\centering
\includegraphics[width=7.5cm, height=5.5cm]{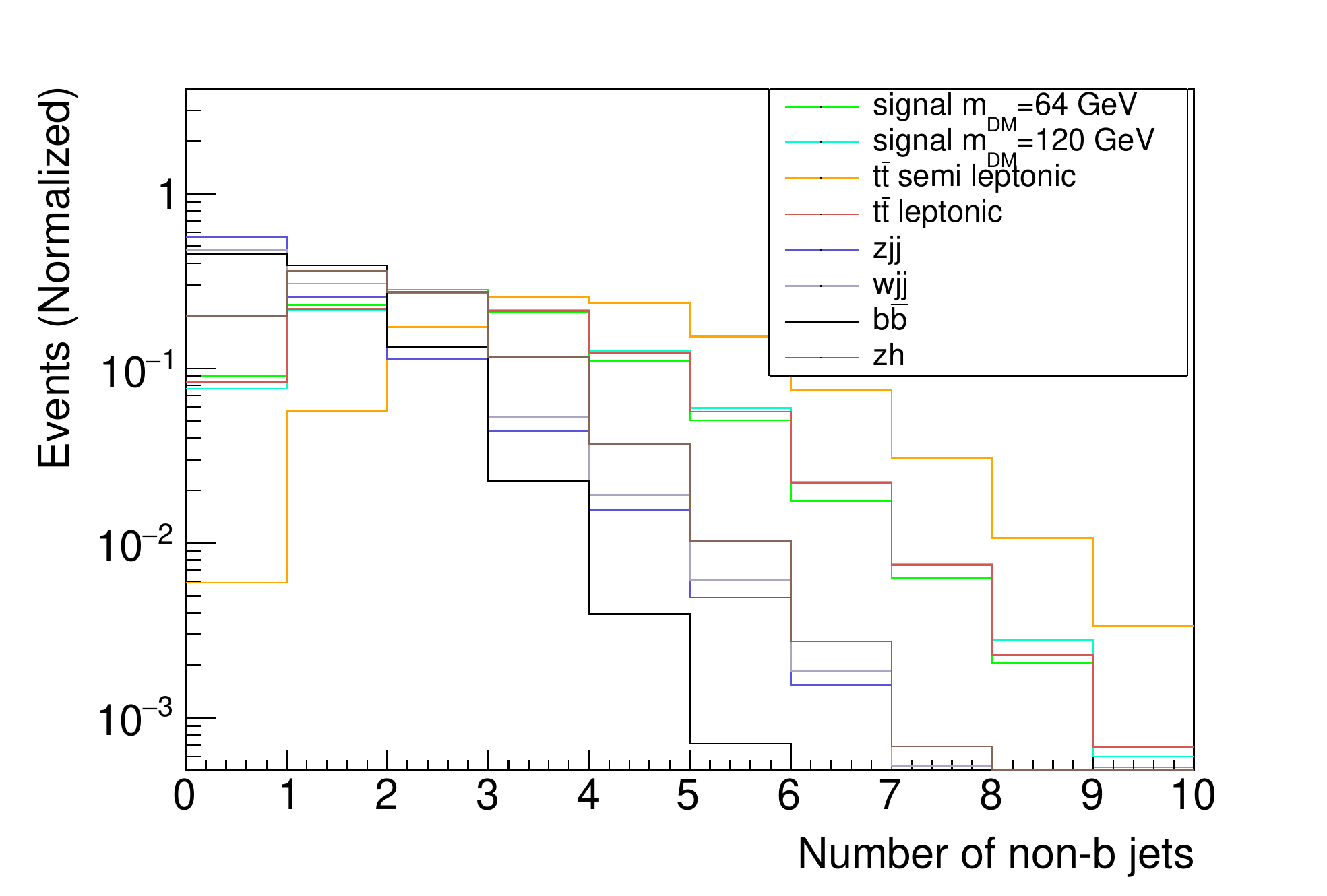} 
\includegraphics[width=7.5cm, height=5.5cm]{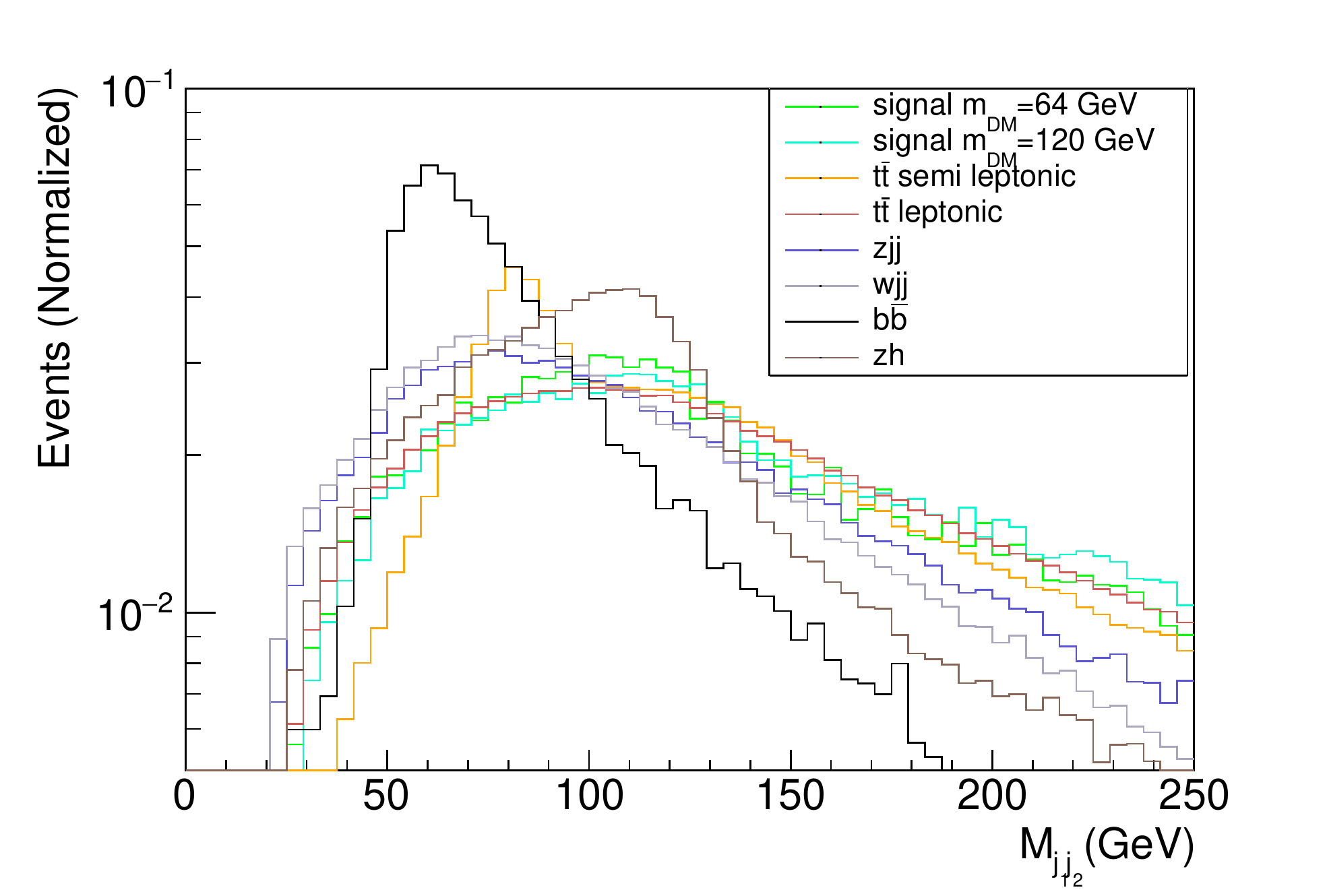}
\caption{jet multiplicity distribution~(left) and invariant mass distribution of leading and sub-leading~(right) light jets. }
\label{jtno_invjj}
\end{figure}

In Figure~\ref{jtno_invjj} (left) we show the jet multiplicity distribution of the signal and background processes. Jet multiplicity distribution here indicates the number of light jets in the process. We know that in $t \bar t$ semileptonic case, which is one of the primary backgrounds, the number of light jets is expected to be more than the signal and other backgrounds, because it has two hard light jets coming from the $W$ decay. This feature can be used to distinguish this background from signal. We also present the distribution of the invariant mass distribution of the leading and sub-leading light jet pair in Figure~\ref{jtno_invjj} (right). In case of $t \bar t$ semileptonic background, the two leading light jets come from $W$ decay and therefore this $m_{jj}$ distribution peaks at $W$ mass. An exclusion of the $m_{jj} \approx m_W$ region in the cut-based analysis as well as a suitable cut on the number of light jets help us control the severe $t \bar t$ background.

\begin{figure}[!hptb]
\centering
\includegraphics[width=7.5cm, height=5.5cm]{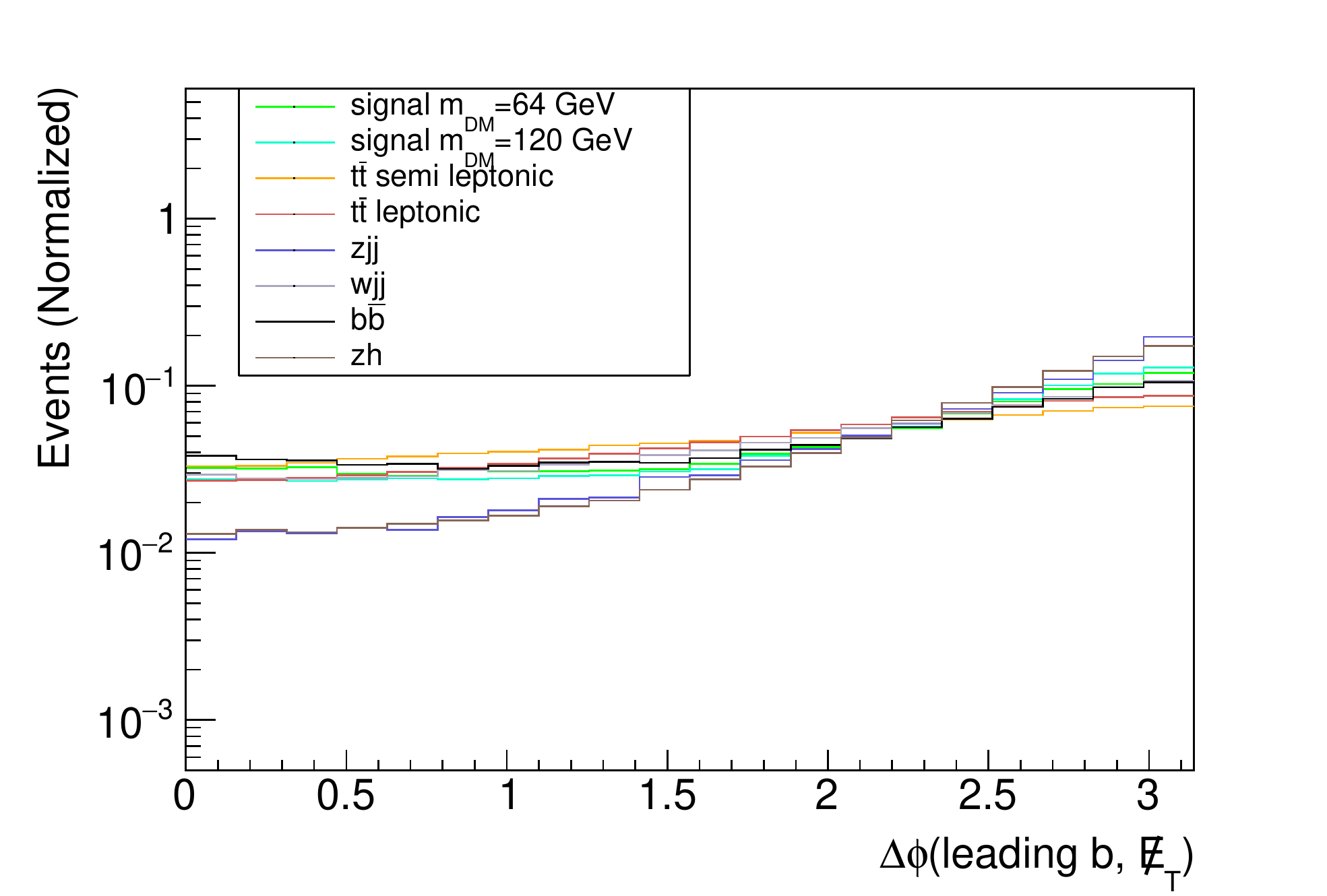} 
\includegraphics[width=7.5cm, height=5.5cm]{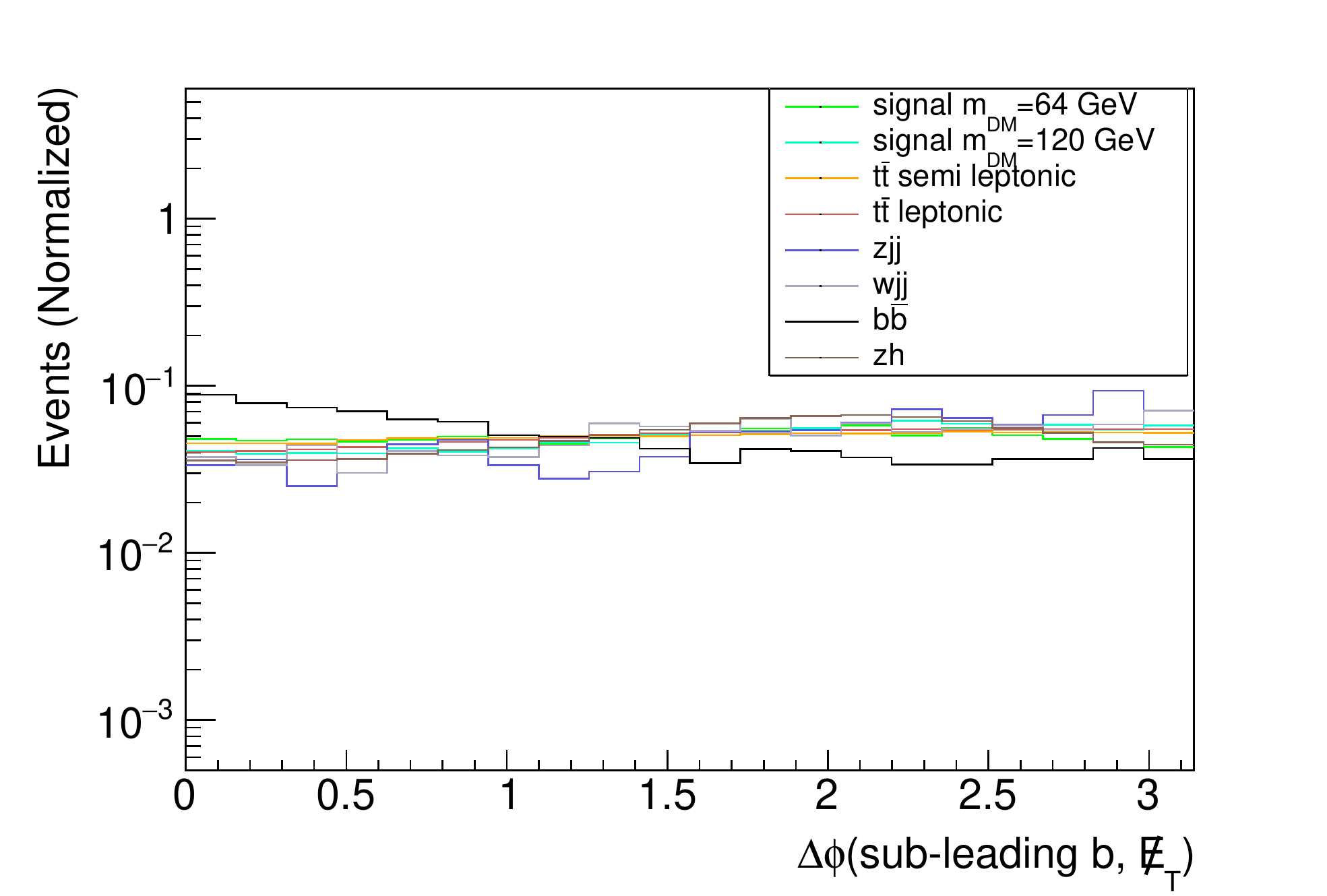}
\caption{$\Delta \phi$ distribution between the $\slashed{E_T}$ and leading~(left) and sub-leading~(right) $b$-jets. }
\label{dphibmet}
\end{figure}

In Figure~\ref{dphibmet} (left) and (right), we plot the $\Delta \phi$ distribution between the $\slashed{E_T}$ and the leading and sub-leading $b$-jets respectively. In case of signal and the $Zh$ background the $b \bar b$ system is exactly opposite to the dark matter system. Therefore the $b$-jets from the Higgs boson decay tend to be mostly opposite to the $\slashed{E_T}$. On the other hand in case of QCD $b \bar b$  events, the the $\slashed{E_T}$ is aligned with either of the $b$-jets because the $\slashed{E_T}$ arises mainly due to the mismeasurement of the $b$-jet energy in this case. This behaviour is visible from the Figure~\ref{dphibmet} (left) and (right).

\bigskip

\paragraph{Results:}

\bigskip

From the discussion on various kinematical observables it is clear that we can choose suitable kinematical cuts on them to enhance the signal-background separation. The following cuts are applied for our analysis.

\begin{itemize}
\item Cut 1: $p_T$ of the leading $b$-jet $> 50$ GeV and  $p_T$ of the sub-leading $b$-jet $> 30$ GeV 
\item Cut 2: $\slashed{E_T} > 200$ GeV
\item Cut 3: $80$ GeV $< M_{b_1 b_2} < 140$ GeV
\item Cut 4: $\Delta \phi(\text{leading}~b, \slashed{E_T}) > 0.35$ and  $\Delta \phi(\text{sub-leading}~b, \slashed{E_T}) > 0.35$ 
\item Cut 5: Number of light jets (not $b$-tagged) $ < 3$
\item Cut 6: Invariant mass of two leading light jet pair $<70$ GeV or $>90$ GeV
\item Cut 7: $\Delta R$ between the leading $b$-jet and the leading light jet $> 1.5$
\end{itemize}

We have applied these cuts on signal and background processes in succession. The cut efficiencies of various cuts for signal and backgrounds are given below. The cut efficiencies quoted are over and above the selection criteria ie. lepton-veto and requirement of two $b$-tagged jets with $p_T> 20 $ GeV.

\begin{table}[!hptb]
\scriptsize{
\begin{tabular}{| c | c | c | c | c | c | c | c | c |}
\hline
Datasets & Xsec(pb) & Cut1 & Cut2 & Cut3 & Cut4 & Cut5 & Cut6 & Cut7   \\
\hline
BP4 & 0.027 & 27.9\% & 7.0\% & 5.0\% & 4.0\% & 2.2\% & 2.1\% & 2.0\% \\
\hline
BP5 & 0.118  & 25.7\% & 3.6\% & 2.3\% & 1.6\% & 1.0\% & 0.9\% & 0.8\% \\
\hline
BP6 & 0.03 & 25.7\% & 3.6\% & 2.3\% & 1.6\% & 1.0\% & 0.9\% & 0.8\% \\
\hline
$t \bar t$ semileptonic & 360 & 14.5\% & 0.3\% & 0.06\% & 0.05\% & 0.016\% & 0.014\% & 0.008\% \\ 
\hline
$t \bar t$ leptonic & 32  & 4.1\% & 0.1\% & 0.02\% & 0.01\% & 0.004\% & 0.0037\% & 0.002\%  \\
\hline
$Z+$jets & 1.7$\times 10^4$ & 0.04\% & 0.002\% & 4.0$\times 10^{-4}$\% & 4.0$\times 10^{-4}$\% & 3.0$\times 10^{-4}$\% & 3.0$\times 10^{-4}$\% & 2.0$\times 10^{-4}$\%  \\
\hline
$W+$jets & 2.4$\times 10^4$ & 0.01\% & 8.1$\times 10^{-5}$\%  & 1.0$\times 10^{-5}$\% & 1.0$\times 10^{-5}$\% & 2.7$\times 10^{-6}$\% & 2.7$\times 10^{-6}$\% & 2.2$\times 10^{-6}$\% \\
\hline
QCD $b \bar b$ &  $10^5$ & 27.3\% & 3.0$\times 10^{-5}$\% & 3.0$\times 10^{-5}$\% & 2.0$\times 10^{-5}$\% & 2.0$\times 10^{-5}$\% & 2.0$\times 10^{-5}$\%  & 1.0$\times 10^{-5}$\%\\
\hline
$Zh$ & 0.07 & 18.9\% & 1.6\% & 1.3\% & 1.2\% & 1.1\% & 1.1\% & 0.09\% \\
\hline
\end{tabular}
\caption{Signal and background efficiencies in the $b \bar b + \slashed{E_T}$ final state after applying various cuts at 13 TeV. The cross-sections are calculated at NLO.}
\label{tablecutflowbb}
}
\end{table}

From the cut-flow efficiencies quoted in Table~\ref{tablecutflowbb} we can see that Cut2 ie. the $\slashed{E_T}$ cut is most essential in eliminating the major backgrounds such as $t \bar t$ semileptonic and QCD $b \bar b$. The other severe background $V+$jets is under control after applying the $b$-veto at the selection level.
Having optimized cut values, we calculate the projected significance (${\cal S})$ for each benchmark point for the 13 TeV LHC with 3000 $fb^{-1}$ in Table~\ref{significancebb}. The formula used for calculating signal significance is given in Eq.~\ref{significance}.

\begin{table}[!hptb]
\begin{center}
\begin{tabular}{| c | c |}
\hline
BP & $
{\cal S}$   \\
\hline
BP4  & 3.6 $\sigma$  \\
\hline
BP5  & 7.0 $\sigma$  \\
\hline
BP6  & 1.6 $\sigma$  \\
\hline
\end{tabular}
\caption{Signal significance for the benchmark points at 13 TeV with ${\cal L}$ = 3000 $fb^{-1}$ in the $b \bar b + \slashed{E_T}$ channel, after using the cuts listed in Table~\ref{tablecutflowbb}. }
\label{significancebb}
\end{center}
\end{table}

From Table~\ref{significancebb} one can see that it is possible to achieve 7$\sigma$ signal significance for BP5 with the $b \bar b +\slashed{E_T}$ final state. As we have chosen BP5 to be exactly same as BP2 of the di-photon case, one can compare the reach of the two channels with this benchmark. We can see that $b \bar b +\slashed{E_T}$ clearly performs better than $\gamma \gamma +\slashed{E_T}$ channel in this regard. BP4 for which dark matter mass is 120 GeV, performs fairly well even within the cut-based framework. 
We have used mainly the strong classifiers in our cut based analysis. However, it is well known in the machine learning literature that a large number of weak classifiers can also lead to a good classification scheme. To that end, we include a range of kinematical variables (weak classifiers) along with our strong classifiers to help us improve our reach in couplings and masses of the dark matter.

\section{Improved analysis through machine learning}\label{sec6}

Having performed the rectangular cut-based analysis in the $\gamma \gamma$ + $\slashed{E_T}$ and $b \bar b + \slashed{E_T}$ channel, we found  that it is possible to achieve considerable signal significance at the HL-LHC for certain regions of the parameter space. Those regions are highly likely to be detected in the future runs. However, there are some benchmarks, namely the ones with small quartic coupling $\lambda_{\Phi\chi}$ which predict rather poor signal significance in a cut-based analysis. We will explore the possibility of probing those regions of parameter space with higher significance with machine learning. As we discussed in the previous section, we go beyond the rectangular cut-based approach here and use more observables, even the weaker classifiers and take into account the correlation between the observables.

In view of this, we present our analysis using packages based on machine learning (ML) techniques. We have performed the analysis with Artificial Neural Network~\cite{Teodorescu:2008zzb} as well as Boosted decision Tree~\cite{Roe:2004na} in order to make a comparison between the two as well as with the cut-based analysis. 
The usefulness of ANN has been widely demonstrated in~\cite{Baldi:2014kfa,Ghosh:2018gyw,Woodruff:2017geg,Oyulmaz:2019jqr,Bhattacherjee:2019fpt} including studies in the Higgs sector~\cite{Hultqvist:1995ibm,Field:1996rw,Bakhet:2015uca,Dey:2019lyr,Lasocha:2020ctd}. For ANN, we have used the toolkit Keras~\cite{keras} with Tensorflow as backend~\cite{tensorflow2015}. For BDT, we have used the package TMVA~\cite{Hocker:2007ht}.
In our present analysis, the $\gamma \gamma$ + $\slashed{E_T}$ channel suffers from low significance due to extremely small signal cross-section. The $b \bar b + \slashed{E_T}$ channel does not suffer from such small signal yield and therefore gives better prospect at discovery compared to $\gamma \gamma$ final state. Like the cut-based analysis here too, we present a comparative study of the performance of these two channels using ML.

\subsection{$\gamma \gamma +\slashed{E_T}$ channel}


For our analysis in the $\gamma \gamma +\slashed{E_T}$ channel, we have used 16 observables as feature variables. Those observables are listed in Table~\ref{featurevargamgam}. An ANN has been constructed feeding these 16 variables in the input layer followed  by 4 hidden layers with nodes 200, 150, 100 and 50 in them respectively. We have used rectified linear unit (RELU) as the activation function acting on the output of each layer. A regularization of 20\% has been applied using dropout. Finally there is a fully connected output layer with binary mode owing to softmax activation function. Categorical cross-entropy was chosen as the loss function with adam as the optimizer~\cite{Kingma:2014vow} for network training with a batch-size 1000 for each epoch, and 100 such epochs. for training we use 80\% of the data, while rest 20\% was kept aside for test or validation of the algorithm. 

\begin{table}[htpb!]
\centering
 \begin{tabular}{||c | c||} 
 \hline
 Variable & Definition \\ [0.5ex] 
 \hline\hline
 $P^{\gamma_1}_{T}$ & Transverse momentum of the leading photon \\ 
 $P^{\gamma_2}_{T}$ & Transverse momentum of the sub-leading photon \\
 $E^{miss}_{T}$ & Missing transverse momentum \\
 $m_{\gamma \gamma}$ & Invariant mass of the leading and sub-leading photons \\
 $\Delta R_{\gamma \gamma}$ & $\Delta R$ between two photons \\
 $N_{j}$ & Number of jets in the event \\
 $\Delta \phi_{\gamma_1 \slashed{E_T}}$ & Azimuthal angle separation between the leading photon and $\slashed{E_T}$ \\
 $\Delta \phi_{\gamma_2 \slashed{E_T}}$ & Azimuthal angle separation between the sub-leading photon and $\slashed{E_T}$ \\
 $\Delta \phi_{j_1 \slashed{E_T}}$ & Azimuthal angle separation between the leading jet and $\slashed{E_T}$ \\
 $\Delta \phi_{j_2 \slashed{E_T}}$ & Azimuthal angle separation between the sub-leading jet and $\slashed{E_T}$ \\
 \hline
 $\eta_{\gamma_1}$ & pseudo-rapidity of the leading photon \\ 
 $\eta_{\gamma_2}$ & pseudo-rapidity of the sub-leading photon \\
 $\Delta \phi_{\gamma \gamma}$ & Azimuthal angle separation between two photons \\
 $M_R$ & Razor variable $M_R$\\
 $M^T_R$ & Razor variable $M^T_R$ \\
 $R$ & Razor variable $R$ \\ [1ex] 
 \hline
 \end{tabular}
 \caption{Feature variables for training in the ML analysis for the $\gamma \gamma +\slashed{E_T}$ channel. The observables which were used in the cut-based analysis have been separated from the new ones by a horizontal line.}
  \label{featurevargamgam}
\end{table}

We introduce a few new observables compared to the cut-based analysis, namely $M_R$, $M^T_R$, $R$~\cite{Chatrchyan:2014goa} among others. These variables are collectively called the Razor variables. The definitions are as follows.

\begin{eqnarray}
M_R &=& \sqrt{(E_{\gamma_1} + E_{\gamma_2})^2 + ({p_Z}_{\gamma_1} + {p_Z}_{\gamma_2})^2}  \\
M^T_R &=& \slashed{E_T} (p_T^{\gamma_1} + p_T^{\gamma_2}) - \slashed{E_T} p_T^{\gamma_1} \cos(\Delta \phi_{\gamma_1 \slashed{E_T}}) -  \slashed{E_T} p_T^{\gamma_2} \cos(\Delta \phi_{\gamma_2 \slashed{E_T}}) \\
R &=& \frac{M^T_R}{M_R}
\end{eqnarray}

\begin{figure}[!hptb]
\centering
\includegraphics[width=8.5cm, height=6.5cm]{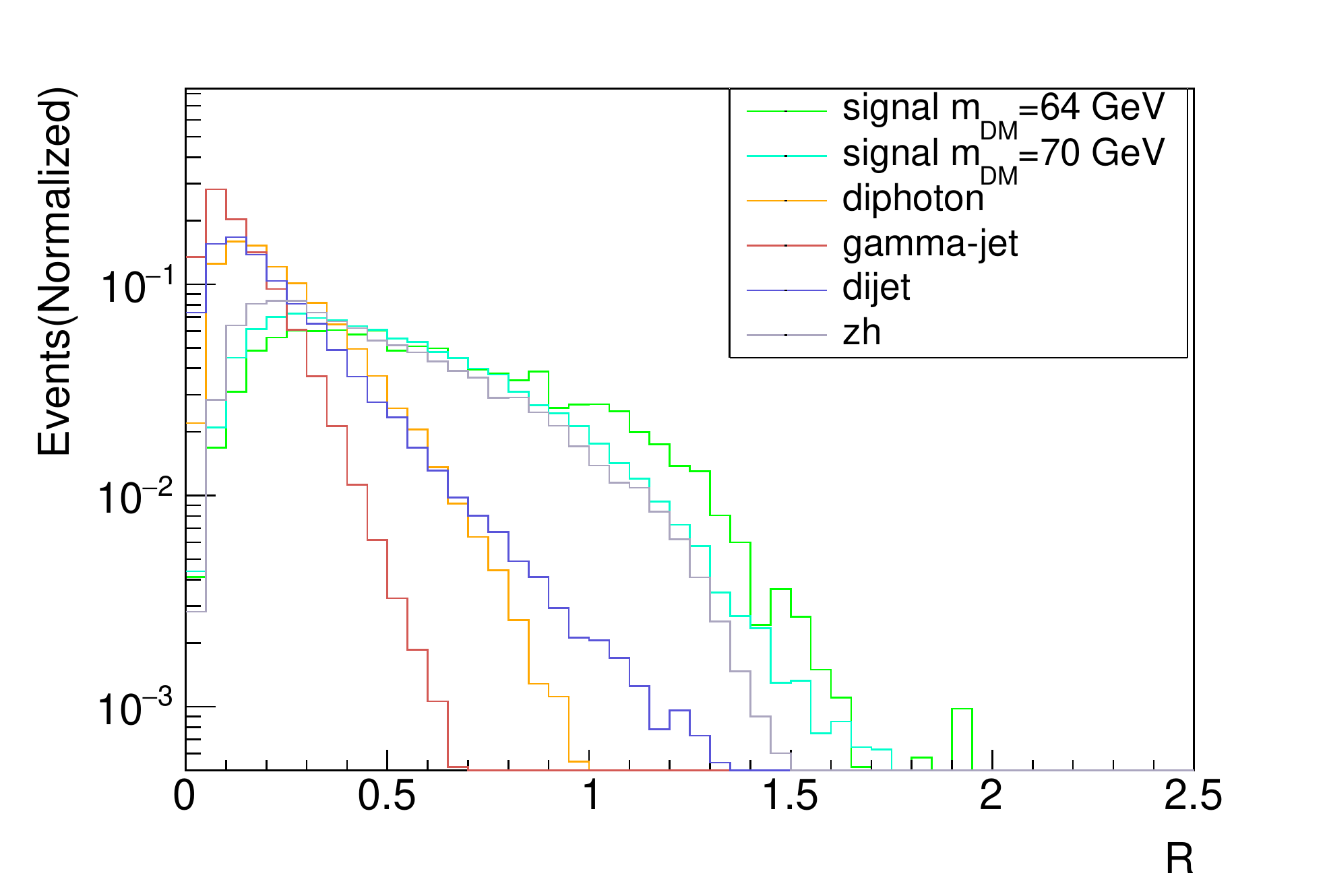} 
\caption{The distribution of Razor variable $R$ for signal and backgrounds. }
\label{razor}
\end{figure}

The $M_R$ variable gives an estimate of the mass scale, which in the limit of massless decay products equals the mass of the parent particle. This variable contains both longitudinal and transverse information. $M^T_R$ on the other hand, is derived only from the transverse momenta of the visible final states and $\slashed{E_T}$. The ratio $R$ between $M_R$ and $M^T_R$ captures the flow of energy along the plane perpendicular to the beam and separating the visible and missing momenta.
We show the distribution of $R$ for signal and background processes in Figure~\ref{razor}, which indicates the variable $R$ and correspondingly all the razor variables possess substantial discriminating power.

\begin{figure}[!hptb]
\centering
\includegraphics[width=7.5cm, height=5.5cm]{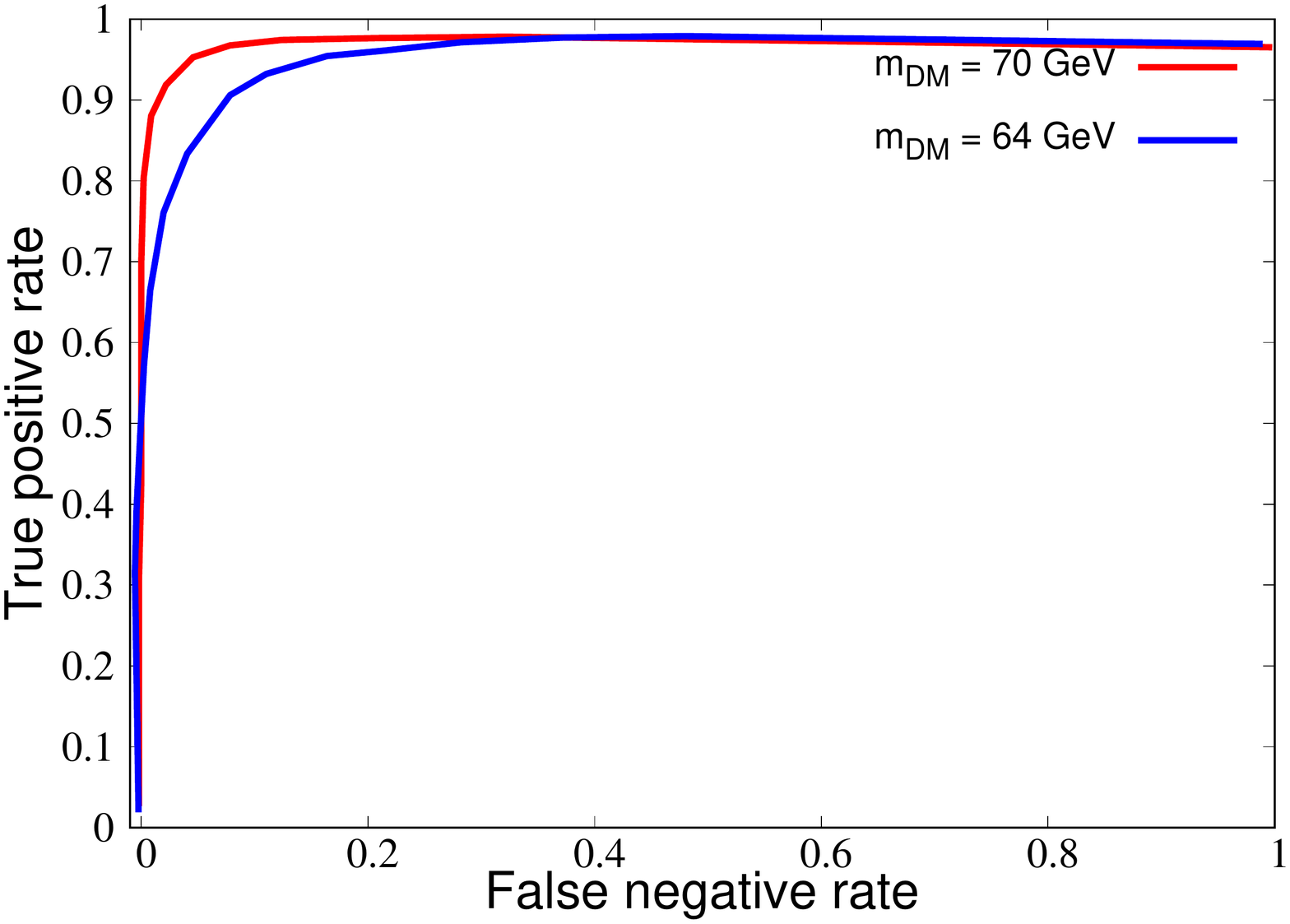} 
\includegraphics[width=7.5cm, height=5.5cm]{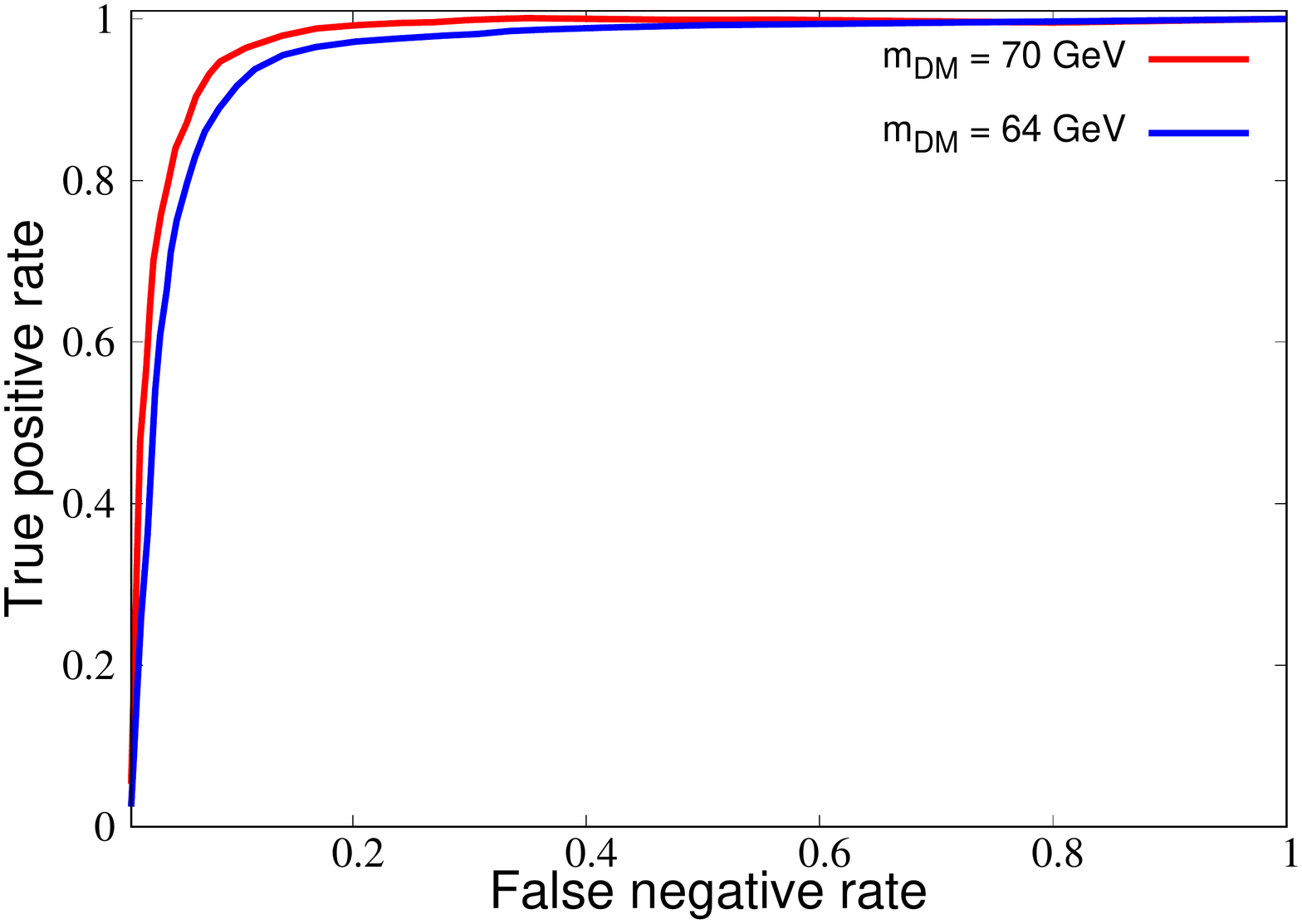} 
\caption{ROC curves for two Dark matter masses in the $\gamma \gamma +\slashed{E_T}$ final state with ANN(left) and BDT(right). } 
\label{roc_gamgam}
\end{figure}

From the BDT analysis, we found out that the $\slashed{E_T}$ and $m_{\gamma \gamma}$ play the most important role in distinguishing between signal and background, which was already expected from our cut-based analysis. $\Delta R_{\gamma \gamma}$, $p_T$ of the leading and sub-leading photons and the Razor variables are also good discriminating variables in this regard. We found that the razor variables (particularly $M^T_R$) are correlated with $\slashed{E_T}$. One should also note that the significant background rejection (di-jet, $\gamma+$ jet) happens while two isolated photons are demanded as has been discussed in the cut-based analysis. 

In Figure~\ref{roc_gamgam}, we show the Receiver Operating Characteristic (ROC)  curves for two mass points $m_{\chi} = 64$ GeV and $m_{\chi} = 70$ GeV from the ANN(left) and BDT(right) analyses. We can see from Figure~\ref{roc_gamgam}, that with increase in the dark matter mass, the discriminating power increases, the ROC curve for $m_{\chi} = 70$ GeV performs better than $m_{\chi} = 64$ GeV. The area under the ROC curve is 0.994(0.993) for $m_{\chi} = 70$ GeV and 0.992(0.991) for $m_{\chi} = 64$ GeV using ANN(BDT). In Table~\ref{significancegamgamann}, we present the signal significance for BP1, BP2 and BP3 (see Table~\ref{bpgam}) using both ANN and BDT. We have scanned along the ROC curves and presented results for selected true positive and false negative rates that will yield the best signal significance for respective benchmark points. The signal significance is calculated using the formula~\ref{significance}.

 \begin{table}[!hptb]
\begin{center}
\begin{tabular}{| c | c | c |}
\hline
BP & ${\cal S}$ (BDT) & ${\cal S}$ (ANN)    \\
\hline
BP1 & 5.3 $\sigma$ & 6.6 $\sigma$  \\
\hline
BP2 & 6.5 $\sigma$ & 7.5 $\sigma$  \\
\hline
BP3 & 2.8 $\sigma$ & 3.3 $\sigma$  \\
\hline
\end{tabular}
\caption{Signal significance for the benchmark points at 13 TeV with ${\cal L}$ = 3000 $fb^{-1}$ in the $\gamma \gamma +\slashed{E_T}$ final state, using ML. }
\label{significancegamgamann}
\end{center}
\end{table}

Comparing Table~\ref{significancegamgamann} with Table~\ref{significancegamgam}, it is clear that machine learning improves the results of our cut-based analysis to a large extent. We also see that ANN performs better than BDT in this case. With machine learning techniques at our disposal, regions with weaker dark matter couplings can be probed at the HL-LHC in the $\gamma \gamma +\slashed{E_T}$ final state, which was unattainable through an exclusive rectangular cut-based method.

\subsection{$b \bar b +\slashed{E_T}$ channel}


We proceed towards the analysis in $b \bar b + \slashed{E_T}$ channel with ML in this subsection. Here we have used 22 observables as feature variables which are listed in Table~\ref{featurevarbb}. An ANN was constructed with these 22 variables fed into the input layer and then a similar structure of the network has been used as described in the $\gamma \gamma$ analysis. The input layer is followed  by 4 hidden layers with nodes 200, 150, 100 and 50 respectively using rectified linear unit (RELU) as the activation function acted on the outputs of each layer. The final layer is a fully connected binary output layer with softmax as the activation function.  20\% of dropout has been applied for regularization. For the loss function categorical cross-entropy was chosen with adam as the optimizer~\cite{Kingma:2014vow} and the network was trained  with a batch-size 1000 for each epoch, and 100 such epochs. 90\% of the data was used for the training purpose while rest 10\% was used for test or validation of the algorithm.

\begin{table}[htpb!]
\centering
 \begin{tabular}{||c | c||} 
 \hline
 Variable & Definition \\ [0.5ex] 
 \hline\hline
 $P^{b_1}_{T}$ & Transverse momentum of the leading $b$-jet \\ 
 $P^{b_2}_{T}$ & Transverse momentum of the sub-leading $b$-jet \\
 $E^{miss}_{T}$ & Missing transverse energy \\
 $m_{bb}$ & Invariant mass of the $b$-jet pair \\
 $N_{jets}$ & Number of light jets in the event \\
 $m_{jj}$ & Invariant mass of the leading light and sub-leading light-jet \\
 $\Delta \phi_{b_1\slashed{E_T}}$ & Azimuthal angle separation between leading $b$-jet and $\slashed{E_T}$ \\
 $\Delta \phi_{b_2\slashed{E_T}}$ & Azimuthal angle separation between sub-leading $b$-jet and $\slashed{E_T}$ \\
 $\Delta R_{b_1j_1}$ & $\Delta R$ between leading $b$-jet and leading light jet \\
 \hline
 $\Delta R_{bb}$ & $\Delta R$ between two leading $b$-jets \\
 $\Delta \phi_{bb}$ & Azimuthal angle separation between two $b$-jets \\
 $H_T$ & Scalar sum of $p_T$ of all visible final states \\
 $\slashed{E_T}$ significance & $\slashed{E_T}/\sqrt{H_T}$  \\
 $\Delta \phi_{b_1j_1}$ & Azimuthal angle separation between leading $b$-jet and leading light jet \\
 $\Delta \phi_{b_2j_1}$ & Azimuthal angle separation between sub-leading $b$-jet and leading light jet \\
 $\Delta \phi_{b_1j_2}$ & Azimuthal angle separation between leading $b$-jet and sub-leading light jet \\
 $\Delta \phi_{b_2j_2}$ & Azimuthal angle separation between sub-leading $b$-jet and sub-leading light jet \\
 $\Delta R_{b_2j_1}$ & $\Delta R$ between sub-leading $b$-jet and leading light-jet \\
 $\Delta R_{b_1j_2}$ & $\Delta R$ between leading $b$-jet and sub-leading light-jet \\
 $\Delta R_{b_2j_2}$ & $\Delta R$ between sub-leading $b$-jet and sub-leading light jet \\
 $\Delta \phi_{j_1\slashed{E_T}}$ & Azimuthal angle separation between leading light-jet and $\slashed{E_T}$ \\
 $\Delta \phi_{j_2\slashed{E_T}}$ & Azimuthal angle separation between sub-leading light-jet and $\slashed{E_T}$ \\ [1ex] 
 \hline
 \end{tabular}
 \caption{Feature variables for training in the ML analysis for the $b \bar{b} +\slashed{E_T}$ channel. The observables which were used in the cut-based analysis have been separated from the new ones by a horizontal line.}
  \label{featurevarbb}
\end{table}

\begin{figure}[!hptb]
\centering
\includegraphics[width=7.5cm, height=5.5cm]{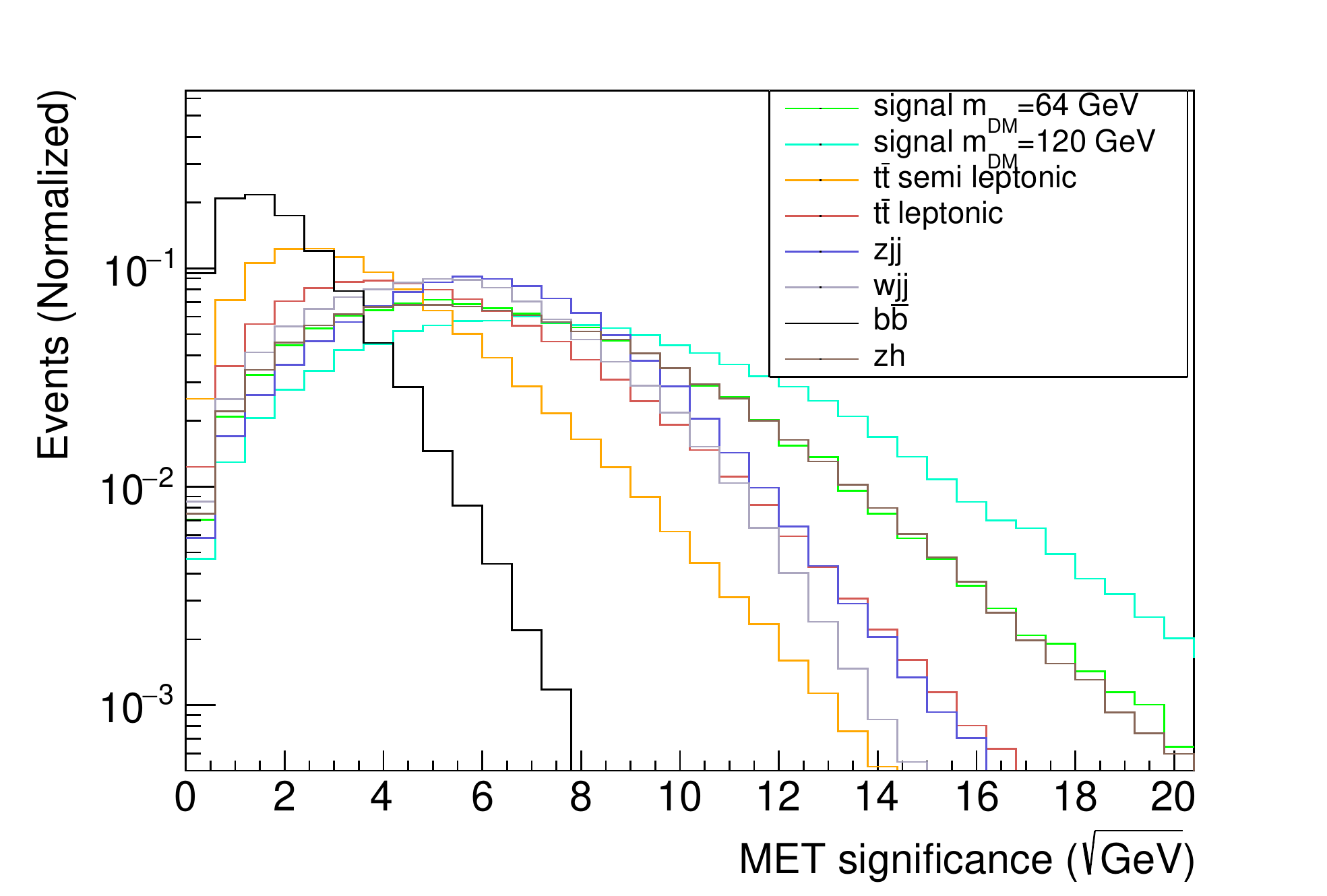} 
\includegraphics[width=7.5cm, height=5.5cm]{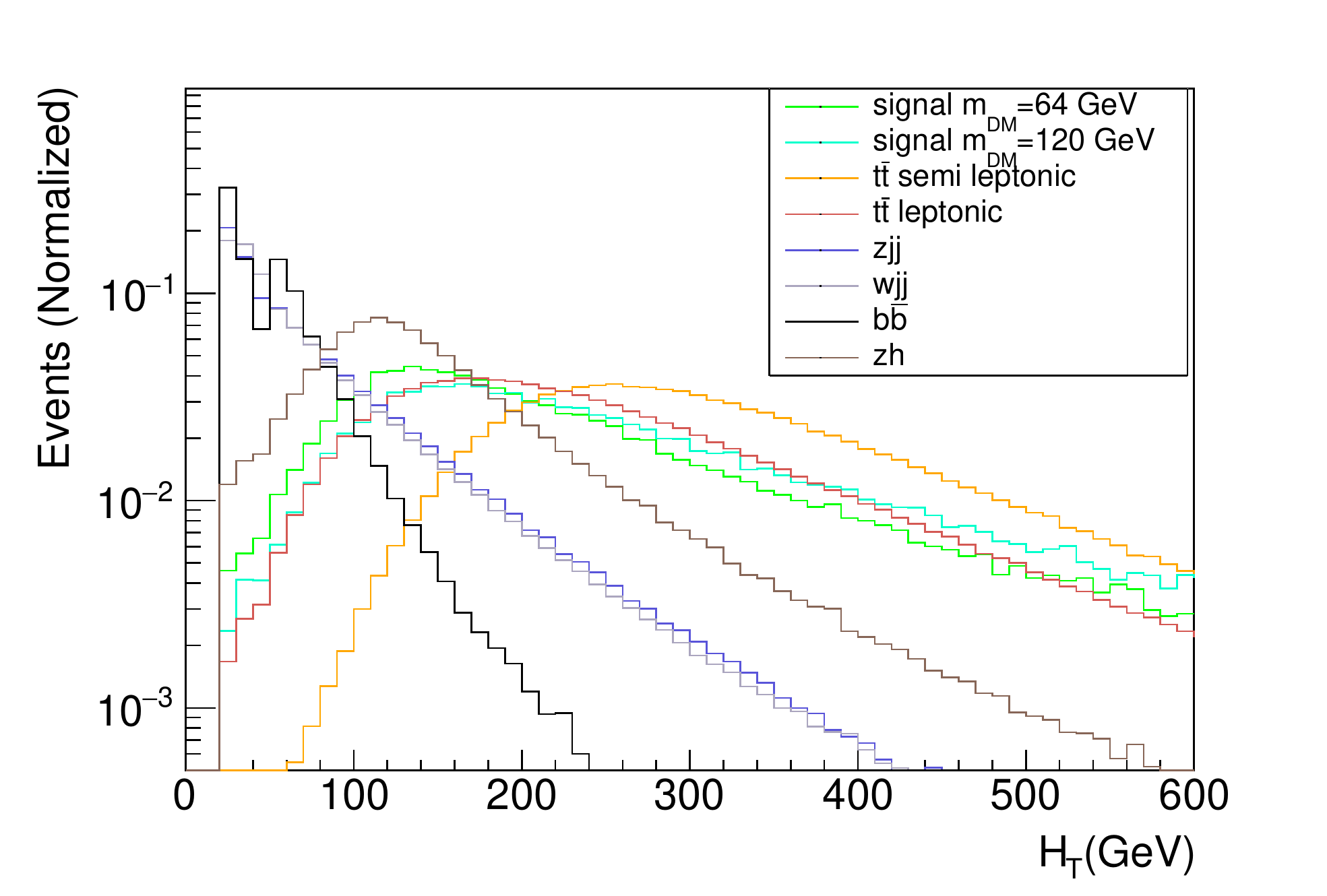} 
\caption{The distribution of $\slashed{E_T}$ significance (left) and $H_T$ (right) for signal and backgrounds. }
\label{metsig}
\end{figure}

We can see from Table~\ref{featurevarbb}, that a number of new observables have been introduced for the ML analysis, compared to the cut-based approach. One such important addition is the $\slashed{E_T}$ significance which has been widely used in experimental analyses for the mono-Higgs + DM search in the $b \bar b + \slashed{E_T}$ final state~\cite{Aaboud:2017yqz}. The $\slashed{E_T}$ significance is defined as the ratio of $\slashed{E_T}$ and the square-root of scalar sum of $p_T$ of all the visible final states ($\sqrt{H_T}$). This observable (albeit correlated with $\slashed{E_T}$) is particularly useful in reducing the $b \bar b$ background as pointed out in ~\cite{Fabiani:2020voj}. We can see the distribution of $\slashed{E_T}$ significance for signal and background from Figure~\ref{metsig} (left). We show the distribution of $H_T$ for signal and backgrounds in Figure~\ref{metsig} (right).

\begin{figure}[!hptb]
\centering
\includegraphics[width=7.5cm, height=5.5cm]{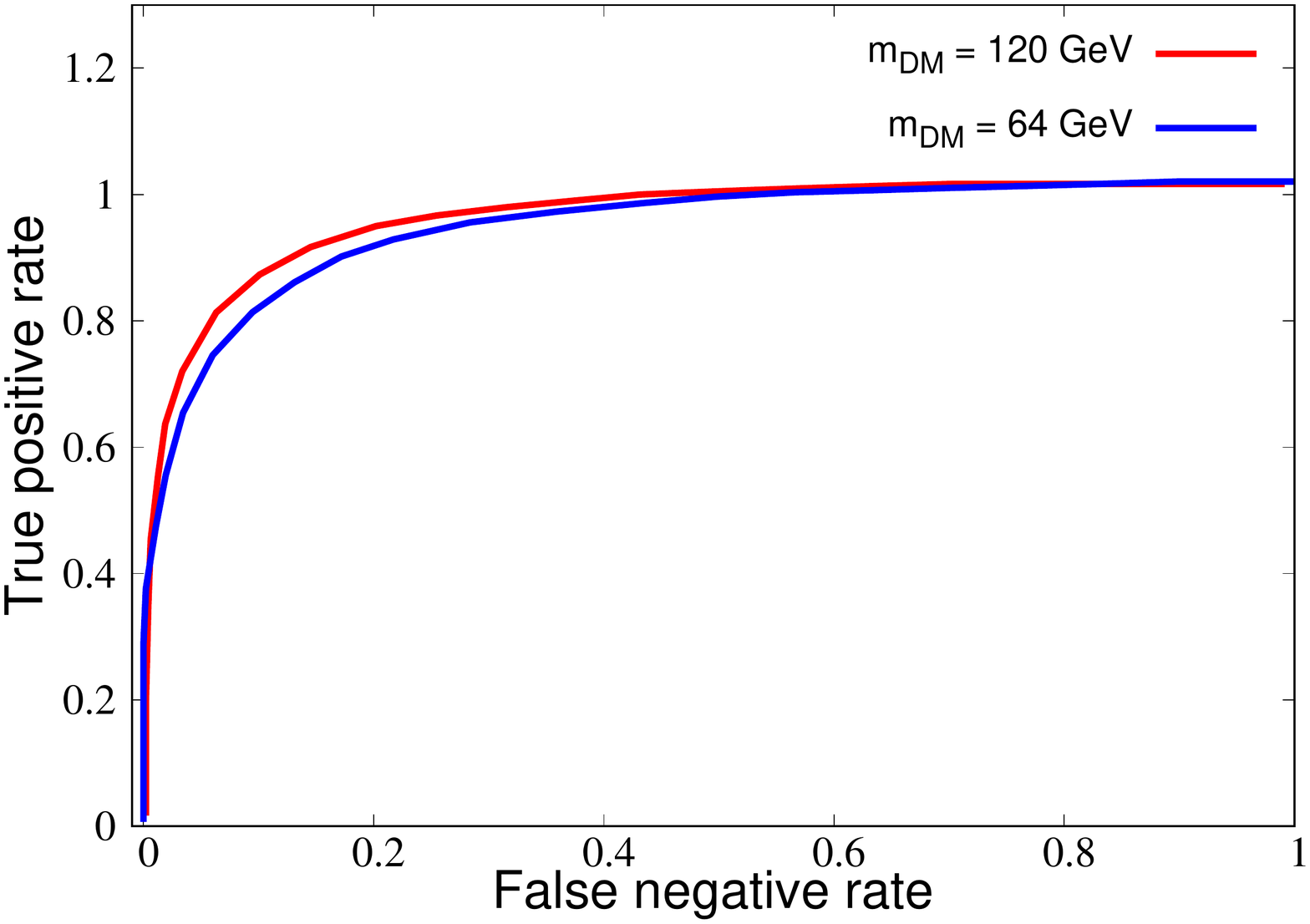} 
\includegraphics[width=7.5cm, height=5.5cm]{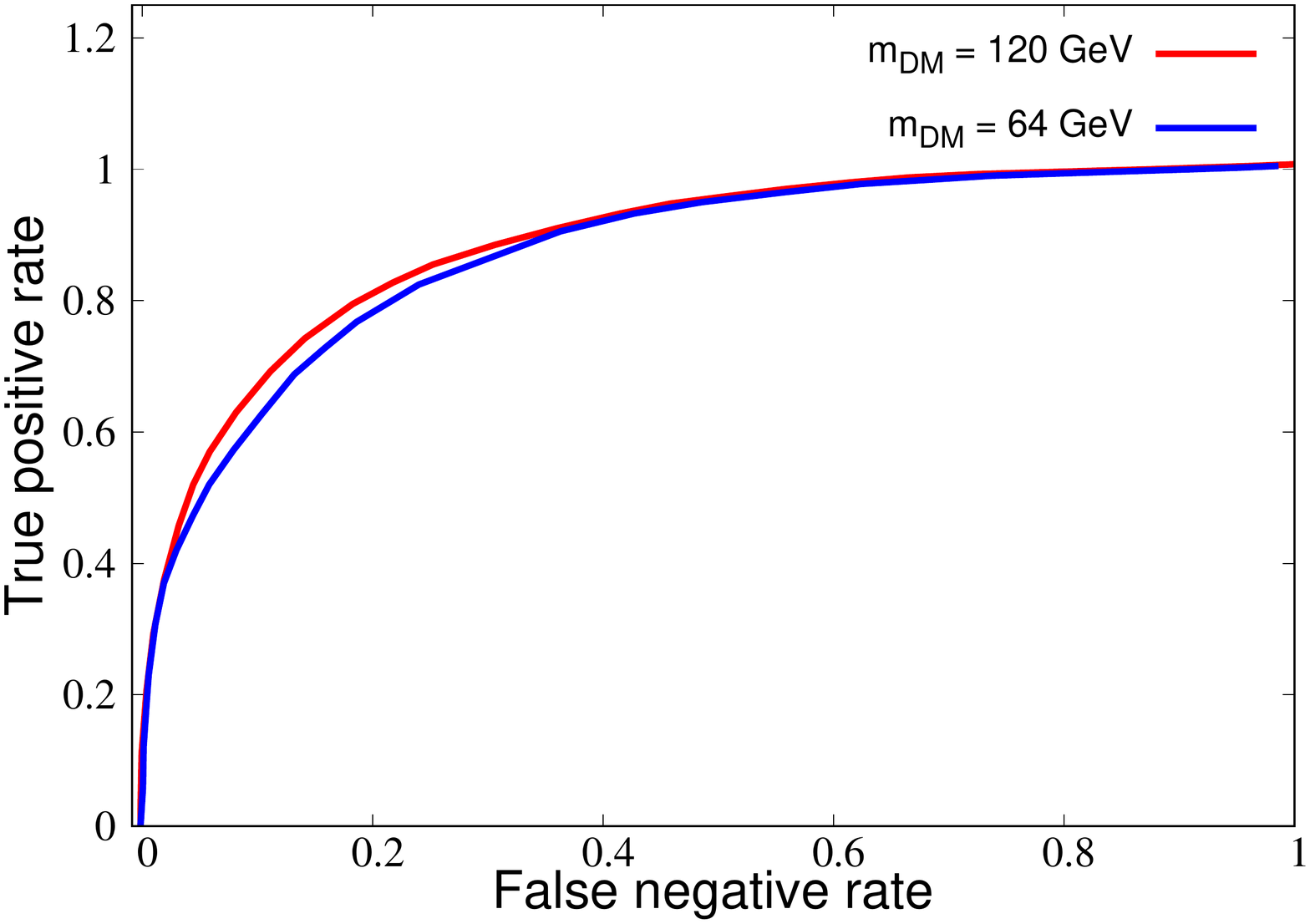} 
\caption{ROC curves for two Dark matter masses in the $b \bar b +\slashed{E_T}$ final state with ANN (left) and BDT(right). }
\label{roc_bbbar}
\end{figure}

BDT analysis ranks $\slashed{E_T}$, $m_{bb}$, $H_T$, $N_{jets}$ observables highest in terms of signal and background separation, reinforcing our understanding from the cut-based analysis. In Figure~\ref{roc_bbbar}, we present the ROC curve for two mass points for the dark matter particle $\chi$, $m_{\chi} = 64$ GeV and $m_{\chi} = 120$ GeV using ANN(left) as well as BDT(right). We can see that as the mass of the dark matter increases, the discriminating power between signal and background increases. This can certainly be attributed to the larger $\slashed{E_T}$ in case of heavier dark matter mass which is clear from Figure~\ref{metinvbb} (left). The area under the ROC curve is 0.95(0.93) in case of $m_{\chi} = 120$ GeV and 0.93(0.91) for $m_{\chi} = 64$ GeV using ANN(BDT). In Table~\ref{significancebbann}, we present the signal significance for the benchmark points given in Table~\ref{bpbb}, calculated using Eq.~\ref{significance}. Here also, we quote the ANN as well as BDT results for the sake of completeness. We would like to mention that we have performed a scan along the ROC curve and chosen the true positive and false negative rates that yield the best signal significance for the respective benchmarks.

\begin{table}[!hptb]
\begin{center}
\begin{tabular}{| c | c | c |}
\hline
BP & ${\cal S}$ (BDT) & ${\cal S}$ (ANN)   \\
\hline
BP4 & 3.9 $\sigma$ (300 $fb^{-1}$) & 4.8 $\sigma$ (300 $fb^{-1}$)  \\
\hline
BP5 & 9.4 $\sigma$ (300 $fb^{-1}$) & 11.5 $\sigma$ (300 $fb^{-1}$) \\
\hline
BP6 & 3.7 $\sigma$  (3000 $fb^{-1}$)& 4.7 $\sigma$  (3000 $fb^{-1}$) \\
\hline
\end{tabular}
\caption{Signal significance for the benchmark points at 13 TeV, in the $b \bar b + \slashed{E_T}$ final state, using ML. }
\label{significancebbann}
\end{center}
\end{table}

A comparison between Table~\ref{significancebb} and Table~\ref{significancebbann} shows that machine learning significantly improves the prospect of the $b \bar b + \slashed{E_T}$ channel. Here too, ANN performs consistently better than BDT. We can see that the large quartic couplings $\lambda_{\Phi \chi}$ (BP 4 and BP 5) can be probed with 300 $fb^{-1}$ luminosity, while lower $\lambda_{\Phi \chi}$ (BP 6) can be probed with $\sim 5 \sigma$ significance with 3000 $fb^{-1}$ luminosity. Furthermore, even smaller values of $\lambda_{\Phi \chi}$ and higher dark matter masses can be probed at the HL-LHC, although the statistical significance may be a little more modest.

\section{Summary and Conclusion}\label{sec7}

In this work, we concentrate on the collider search for dark matter in the mono-Higgs + $\slashed{E_T}$ final state. This channel along with mono-jet, mono-photon, and mono-V final states has garnered substantial interest among experimentalists and theorists alike for dark matter hunting. In the present work, we have chosen the dark matter to be a WIMP-like scalar that interacts with the SM particles via Higgs mediation. This scenario (the so-called Higgs portal mechanism) is severely constrained by the observed relic density of the universe and the direct detection experiments. The part of the parameter space that satisfies direct-detection constraints fails to produce the observed relic and can be rescued only by some additional interactions between the Higgs boson, the SM particles, and the postulated scalar DM. Such interactions can be provided by possible new physics at the TeV scale and can be modelled by higher dimension operators. One chooses the coupling strengths suitable to yield partial cancellation in the direct detection experiments and still reproduce the observed relic density. In addition, the presence of higher-dimensional operators may enhance the production cross-section for the $h+\chi\chi$ state. Furthermore, the detection of mono-Higgs signals with appreciable rates could serve as a pointer to additional
interactions of the dark matter candidate particle, since a Higgs portal scenario with such interaction strength
would otherwise contradict observations. Our postulated dimension-6 interaction terms fit into that role rather well.  

Accounting for the constraints coming from the latest results on the observed relic density, the direct detection, indirect detection, and invisible decay rate of the Higgs boson we choose benchmark points that yield large production cross-section for $h+\chi\chi$ state at the LHC. We would like to point out that the resonant Higgs mass region ($m_\chi \sim m_h/2$) is the best possible region to probe in the mono-Higgs final state because of the large production cross-section. We choose the $\gamma\gamma+\slashed{E_T}$ and $b\bar b +  \slashed{E_T}$ final states, corresponding to the di-photon and $b\bar b$ decay channel of the Higgs, for our study. The di-photon channel is chosen due to smaller background contributions despite a smaller branching ratio while the $b\bar b$ channel is chosen for a large branching ratio despite troublesome background contributions. 

For the di-photon channel, we go beyond the SM backgrounds with prompt photons usually considered in the literature and estimate the fake/non-prompt photons coming from QCD di-jet and $\gamma+$ jets events. We find that the tail of these backgrounds with large production cross-sections can be detrimental for signal background separation in the di-photon channel however strong isolation of photons along with suitable cuts on kinematical observables enable us to get considerable signal significance at the high luminosity (3000  $fb^{-1}$) LHC. For the $b\bar b$ channel we simulate full backgrounds coming from QCD multi-jets, $V+$jets, and $t\bar t$ final states. We find that the $b\bar b$ channel fares better than the di-photon channel owing to larger rates (despite the large background rates). This allows us to probe higher dark matter masses and/or smaller quartic couplings in the $b\bar b$ channel using suitable cuts and relevant observables.

We consider (almost) an exhaustive list of kinematical variables to perform BDT- and ANN-based analysis of the signal significance. We find that both these method leads to a significant improvement in the signal significance for all three benchmark points in the di-photon channel with ANN performing better than BDT. For the $b\bar b$ channel the improvement is even more exciting: Two of the large coupling benchmark points can be probed with $\ge 5\sigma$ significance even with a luminosity of only 300 fb$^{-1}$. Again the ANN performed better than BDT. 

In principle, it is possible to extend this analysis to other Higgs decay modes
such as $WW, ZZ$ and $\tau\tau$. However, such channels are either {\em prima facie}
beset with low  rates for the viable final states, or have challenges in reconstruction of the Higgs
peak. It may be useful to try the $\tau\tau$ mode in particular as a confirmatory channel. It remains a challenge to  see how much improvement occurs via neural network techniques. We plan to take this up in a follow-up study.

\section{Acknowledgements}

We would like to thank Shamik Ghosh for useful discussions and help with the ANN codes. The work of JL and BM
is partially supported by funding available from the Department of Atomic Energy, Government of
India, for the Regional Centre for Accelerator-based Particle Physics (RECAPP), Harish-
Chandra Research Institute. The work of RKS is partially supported by SERB, DST, Government of India through the project EMR/2017/002778. JL would like to thank Saha Institute of Nuclear Physics, HBNI and Indian Institute of Science Education and Research, Kolkata for their hospitality where substantial part of this work was done. SB and DB would like to thank RECAPP, HRI for their hospitality, where initial part of the work was done.

\bibliographystyle{JHEP}
\bibliography{paperbib}

\providecommand{\href}[2]{#2}\begingroup\raggedright\begin{thebibliography}{10}

\bibitem{Abe:1995hr}
{\scshape CDF} collaboration, F.~Abe et~al., \emph{{Observation of top quark
  production in $\bar{p}p$ collisions}},
  \href{https://doi.org/10.1103/PhysRevLett.74.2626}{\emph{Phys. Rev. Lett.}
  {\bfseries 74} (1995) 2626--2631},
  [\href{https://arxiv.org/abs/hep-ex/9503002}{{\ttfamily hep-ex/9503002}}].

\bibitem{D0:1995jca}
{\scshape D0} collaboration, S.~Abachi et~al., \emph{{Observation of the top
  quark}}, \href{https://doi.org/10.1103/PhysRevLett.74.2632}{\emph{Phys. Rev.
  Lett.} {\bfseries 74} (1995) 2632--2637},
  [\href{https://arxiv.org/abs/hep-ex/9503003}{{\ttfamily hep-ex/9503003}}].

\bibitem{Drees:2001xw}
J.~Drees, \emph{{Review of final LEP results, or, A Tribute to LEP}},
  \href{https://doi.org/10.1142/S0217751X02012727}{\emph{Int. J. Mod. Phys.}
  {\bfseries A17} (2002) 3259--3283},
  [\href{https://arxiv.org/abs/hep-ex/0110077}{{\ttfamily hep-ex/0110077}}].

\bibitem{ALEPH:2005ab}
{\scshape ALEPH, DELPHI, L3, OPAL, SLD, LEP Electroweak Working Group, SLD
  Electroweak Group, SLD Heavy Flavour Group} collaboration, S.~Schael et~al.,
  \emph{{Precision electroweak measurements on the $Z$ resonance}},
  \href{https://doi.org/10.1016/j.physrep.2005.12.006}{\emph{Phys. Rept.}
  {\bfseries 427} (2006) 257--454},
  [\href{https://arxiv.org/abs/hep-ex/0509008}{{\ttfamily hep-ex/0509008}}].

\bibitem{Khachatryan:2016vau}
{\scshape ATLAS, CMS} collaboration, G.~Aad et~al., \emph{{Measurements of the
  Higgs boson production and decay rates and constraints on its couplings from
  a combined ATLAS and CMS analysis of the LHC pp collision data at $
  \sqrt{s}=7 $ and 8 TeV}},
  \href{https://doi.org/10.1007/JHEP08(2016)045}{\emph{JHEP} {\bfseries 08}
  (2016) 045}, [\href{https://arxiv.org/abs/1606.02266}{{\ttfamily
  1606.02266}}].

\bibitem{Khachatryan:2014rra}
{\scshape CMS} collaboration, V.~Khachatryan et~al., \emph{{Search for dark
  matter, extra dimensions, and unparticles in monojet events in
  proton–proton collisions at $\sqrt{s} = 8$ TeV}},
  \href{https://doi.org/10.1140/epjc/s10052-015-3451-4}{\emph{Eur. Phys. J.}
  {\bfseries C75} (2015) 235},
  [\href{https://arxiv.org/abs/1408.3583}{{\ttfamily 1408.3583}}].

\bibitem{Diehl:2014dda}
{\scshape ATLAS} collaboration, E.~Diehl, \emph{{The search for dark matter
  using monojets and monophotons with the ATLAS detector}},
  \href{https://doi.org/10.1063/1.4883448}{\emph{AIP Conf. Proc.} {\bfseries
  1604} (2015) 324--330}.

\bibitem{No:2015xqa}
J.~M. No, \emph{{Looking through the pseudoscalar portal into dark matter:
  Novel mono-Higgs and mono-Z signatures at the LHC}},
  \href{https://doi.org/10.1103/PhysRevD.93.031701}{\emph{Phys. Rev.}
  {\bfseries D93} (2016) 031701},
  [\href{https://arxiv.org/abs/1509.01110}{{\ttfamily 1509.01110}}].

\bibitem{Schramm:2016csx}
{\scshape ATLAS} collaboration, S.~Schramm, \emph{{ATLAS Sensitivity to WIMP
  Dark Matter in the Monojet Topology at $\sqrt s=$ 14 TeV}},
  \href{https://doi.org/10.1016/j.nuclphysbps.2015.09.404}{\emph{Nucl. Part.
  Phys. Proc.} {\bfseries 273-275} (2016) 2397--2399}.

\bibitem{Aaboud:2016uro}
{\scshape ATLAS} collaboration, M.~Aaboud et~al., \emph{{Search for new
  phenomena in events with a photon and missing transverse momentum in $pp$
  collisions at $\sqrt{s}=13$ TeV with the ATLAS detector}},
  \href{https://doi.org/10.1007/JHEP06(2016)059}{\emph{JHEP} {\bfseries 06}
  (2016) 059}, [\href{https://arxiv.org/abs/1604.01306}{{\ttfamily
  1604.01306}}].

\bibitem{ATLAS:2018jsf}
{\scshape ATLAS} collaboration, T.~A. collaboration, \emph{{Prospects for Dark
  Matter searches in mono-photon and VBF+$E_T^{miss}$ final states in ATLAS}},
  .

\bibitem{CMS:2018fux}
{\scshape CMS} collaboration, C.~Collaboration, \emph{{Projection of the Mono-Z
  search for dark matter to the HL-LHC}}, .

\bibitem{Ghorbani:2016edw}
K.~Ghorbani and L.~Khalkhali, \emph{{Mono-Higgs signature in a fermionic dark
  matter model}}, \href{https://doi.org/10.1088/1361-6471/aa823a}{\emph{J.
  Phys.} {\bfseries G44} (2017) 105004},
  [\href{https://arxiv.org/abs/1608.04559}{{\ttfamily 1608.04559}}].

\bibitem{Miniello:2017esf}
{\scshape CMS} collaboration, G.~Miniello, \emph{{Searches for Dark Matter via
  Mono-Higgs signatures with the CMS experiment}},
  \href{https://doi.org/10.22323/1.314.0711}{\emph{PoS} {\bfseries EPS-HEP2017}
  (2017) 711}.

\bibitem{Abdallah:2016vcn}
W.~Abdallah, A.~Hammad, S.~Khalil and S.~Moretti, \emph{{Search for Mono-Higgs
  Signals at the LHC in the B-L Supersymmetric Standard Model}},
  \href{https://doi.org/10.1103/PhysRevD.95.055019}{\emph{Phys. Rev.}
  {\bfseries D95} (2017) 055019},
  [\href{https://arxiv.org/abs/1608.07500}{{\ttfamily 1608.07500}}].

\bibitem{Baum:2017gbj}
S.~Baum, K.~Freese, N.~R. Shah and B.~Shakya, \emph{{NMSSM Higgs boson search
  strategies at the LHC and the mono-Higgs signature in particular}},
  \href{https://doi.org/10.1103/PhysRevD.95.115036}{\emph{Phys. Rev.}
  {\bfseries D95} (2017) 115036},
  [\href{https://arxiv.org/abs/1703.07800}{{\ttfamily 1703.07800}}].

\bibitem{Carpenter:2013xra}
L.~Carpenter, A.~DiFranzo, M.~Mulhearn, C.~Shimmin, S.~Tulin and D.~Whiteson,
  \emph{{Mono-Higgs-boson: A new collider probe of dark matter}},
  \href{https://doi.org/10.1103/PhysRevD.89.075017}{\emph{Phys. Rev.}
  {\bfseries D89} (2014) 075017},
  [\href{https://arxiv.org/abs/1312.2592}{{\ttfamily 1312.2592}}].

\bibitem{Petrov:2013nia}
A.~A. Petrov and W.~Shepherd, \emph{{Searching for dark matter at LHC with
  Mono-Higgs production}},
  \href{https://doi.org/10.1016/j.physletb.2014.01.051}{\emph{Phys. Lett.}
  {\bfseries B730} (2014) 178--183},
  [\href{https://arxiv.org/abs/1311.1511}{{\ttfamily 1311.1511}}].

\bibitem{Berlin:2014cfa}
A.~Berlin, T.~Lin and L.-T. Wang, \emph{{Mono-Higgs Detection of Dark Matter at
  the LHC}}, \href{https://doi.org/10.1007/JHEP06(2014)078}{\emph{JHEP}
  {\bfseries 06} (2014) 078},
  [\href{https://arxiv.org/abs/1402.7074}{{\ttfamily 1402.7074}}].

\bibitem{Basso:2015aee}
L.~Basso, \emph{{Resonant mono Higgs at the LHC}},
  \href{https://doi.org/10.1007/JHEP04(2016)087}{\emph{JHEP} {\bfseries 04}
  (2016) 087}, [\href{https://arxiv.org/abs/1512.06381}{{\ttfamily
  1512.06381}}].

\bibitem{Djouadi:2012zc}
A.~Djouadi, A.~Falkowski, Y.~Mambrini and J.~Quevillon, \emph{{Direct Detection
  of Higgs-Portal Dark Matter at the LHC}},
  \href{https://doi.org/10.1140/epjc/s10052-013-2455-1}{\emph{Eur. Phys. J.}
  {\bfseries C73} (2013) 2455},
  [\href{https://arxiv.org/abs/1205.3169}{{\ttfamily 1205.3169}}].

\bibitem{Greljo:2013wja}
A.~Greljo, J.~Julio, J.~F. Kamenik, C.~Smith and J.~Zupan, \emph{{Constraining
  Higgs mediated dark matter interactions}},
  \href{https://doi.org/10.1007/JHEP11(2013)190}{\emph{JHEP} {\bfseries 11}
  (2013) 190}, [\href{https://arxiv.org/abs/1309.3561}{{\ttfamily 1309.3561}}].

\bibitem{Han:2016gyy}
H.~Han, J.~M. Yang, Y.~Zhang and S.~Zheng, \emph{{Collider Signatures of
  Higgs-portal Scalar Dark Matter}},
  \href{https://doi.org/10.1016/j.physletb.2016.03.010}{\emph{Phys. Lett.}
  {\bfseries B756} (2016) 109--112},
  [\href{https://arxiv.org/abs/1601.06232}{{\ttfamily 1601.06232}}].

\bibitem{Aaboud:2019yqu}
{\scshape ATLAS} collaboration, M.~Aaboud et~al., \emph{{Constraints on
  mediator-based dark matter and scalar dark energy models using $\sqrt s = 13$
  TeV $pp$ collision data collected by the ATLAS detector}},
  \href{https://doi.org/10.1007/JHEP05(2019)142}{\emph{JHEP} {\bfseries 05}
  (2019) 142}, [\href{https://arxiv.org/abs/1903.01400}{{\ttfamily
  1903.01400}}].

\bibitem{Arcadi:2019lka}
G.~Arcadi, A.~Djouadi and M.~Raidal, \emph{{Dark Matter through the Higgs
  portal}}, \href{https://doi.org/10.1016/j.physrep.2019.11.003}{\emph{Phys.
  Rept.} {\bfseries 842} (2020) 1--180},
  [\href{https://arxiv.org/abs/1903.03616}{{\ttfamily 1903.03616}}].

\bibitem{Gross:2017dan}
C.~Gross, O.~Lebedev and T.~Toma, \emph{{Cancellation Mechanism for
  Dark-Matter–Nucleon Interaction}},
  \href{https://doi.org/10.1103/PhysRevLett.119.191801}{\emph{Phys. Rev. Lett.}
  {\bfseries 119} (2017) 191801},
  [\href{https://arxiv.org/abs/1708.02253}{{\ttfamily 1708.02253}}].

\bibitem{Dey:2019lyr}
A.~Dey, J.~Lahiri and B.~Mukhopadhyaya, \emph{{LHC signals of a heavy doublet
  Higgs as dark matter portal: cut-based approach and improvement with gradient
  boosting and neural networks}},
  \href{https://doi.org/10.1007/JHEP09(2019)004}{\emph{JHEP} {\bfseries 09}
  (2019) 004}, [\href{https://arxiv.org/abs/1905.02242}{{\ttfamily
  1905.02242}}].

\bibitem{Okada:2020zxo}
N.~Okada, D.~Raut and Q.~Shafi, \emph{{Pseudo-Goldstone Dark Matter in gauged
  $B-L$ extended Standard Model}},
  \href{https://arxiv.org/abs/2001.05910}{{\ttfamily 2001.05910}}.

\bibitem{Ade:2013zuv}
{\scshape Planck} collaboration, P.~A.~R. Ade et~al., \emph{{Planck 2013
  results. XVI. Cosmological parameters}},
  \href{https://doi.org/10.1051/0004-6361/201321591}{\emph{Astron. Astrophys.}
  {\bfseries 571} (2014) A16},
  [\href{https://arxiv.org/abs/1303.5076}{{\ttfamily 1303.5076}}].

\bibitem{Aprile:2018dbl}
{\scshape XENON} collaboration, E.~Aprile et~al., \emph{{Dark Matter Search
  Results from a One Ton-Year Exposure of XENON1T}},
  \href{https://doi.org/10.1103/PhysRevLett.121.111302}{\emph{Phys. Rev. Lett.}
  {\bfseries 121} (2018) 111302},
  [\href{https://arxiv.org/abs/1805.12562}{{\ttfamily 1805.12562}}].

\bibitem{Ackermann:2015zua}
{\scshape Fermi-LAT} collaboration, M.~Ackermann et~al., \emph{{Searching for
  Dark Matter Annihilation from Milky Way Dwarf Spheroidal Galaxies with Six
  Years of Fermi Large Area Telescope Data}},
  \href{https://doi.org/10.1103/PhysRevLett.115.231301}{\emph{Phys. Rev. Lett.}
  {\bfseries 115} (2015) 231301},
  [\href{https://arxiv.org/abs/1503.02641}{{\ttfamily 1503.02641}}].

\bibitem{Ahnen:2016qkx}
{\scshape MAGIC, Fermi-LAT} collaboration, M.~L. Ahnen et~al., \emph{{Limits to
  Dark Matter Annihilation Cross-Section from a Combined Analysis of MAGIC and
  Fermi-LAT Observations of Dwarf Satellite Galaxies}},
  \href{https://doi.org/10.1088/1475-7516/2016/02/039}{\emph{JCAP} {\bfseries
  1602} (2016) 039}, [\href{https://arxiv.org/abs/1601.06590}{{\ttfamily
  1601.06590}}].

\bibitem{Sirunyan:2018owy}
{\scshape CMS} collaboration, A.~M. Sirunyan et~al., \emph{{Search for
  invisible decays of a Higgs boson produced through vector boson fusion in
  proton-proton collisions at $\sqrt{s} =$ 13 TeV}},
  \href{https://doi.org/10.1016/j.physletb.2019.04.025}{\emph{Phys. Lett.}
  {\bfseries B793} (2019) 520--551},
  [\href{https://arxiv.org/abs/1809.05937}{{\ttfamily 1809.05937}}].

\bibitem{CMS:2016xok}
{\scshape CMS} collaboration, C.~Collaboration, \emph{{Search for Dark Matter
  Produced in Association with a Higgs Boson Decaying to Two Photons}}, .

\bibitem{CMS-PAS-EXO-16-054}
{\scshape CMS Collaboration} collaboration, \emph{{Search for Dark Matter
  Produced in Association with a Higgs Boson Decaying to Two Photons}},  Tech.
  Rep. CMS-PAS-EXO-16-054, CERN, Geneva, 2017.

\bibitem{Sirunyan:2017hnk}
{\scshape CMS} collaboration, A.~M. Sirunyan et~al., \emph{{Search for
  associated production of dark matter with a Higgs boson decaying to $
  \mathrm{b}\overline{\mathrm{b}} $ or $\gamma \gamma$ at $ \sqrt{s}=13$ TeV}},
  \href{https://doi.org/10.1007/JHEP10(2017)180}{\emph{JHEP} {\bfseries 10}
  (2017) 180}, [\href{https://arxiv.org/abs/1703.05236}{{\ttfamily
  1703.05236}}].

\bibitem{Sirunyan:2018fpy}
{\scshape CMS} collaboration, A.~M. Sirunyan et~al., \emph{{Search for dark
  matter produced in association with a Higgs boson decaying to $\gamma\gamma$
  or $\tau^+\tau^-$ at $\sqrt{s} =$ 13 TeV}},
  \href{https://doi.org/10.1007/JHEP09(2018)046}{\emph{JHEP} {\bfseries 09}
  (2018) 046}, [\href{https://arxiv.org/abs/1806.04771}{{\ttfamily
  1806.04771}}].

\bibitem{Sirunyan:2019zav}
{\scshape CMS} collaboration, A.~M. Sirunyan et~al., \emph{{Search for dark
  matter particles produced in association with a Higgs boson in proton-proton
  collisions at $ \sqrt{\mathrm{s}} $ = 13 TeV}},
  \href{https://doi.org/10.1007/JHEP03(2020)025}{\emph{JHEP} {\bfseries 03}
  (2020) 025}, [\href{https://arxiv.org/abs/1908.01713}{{\ttfamily
  1908.01713}}].

\bibitem{Aad:2015yga}
{\scshape ATLAS} collaboration, G.~Aad et~al., \emph{{Search for Dark Matter in
  Events with Missing Transverse Momentum and a Higgs Boson Decaying to Two
  Photons in $pp$ Collisions at $\sqrt{s}=8$ TeV with the ATLAS Detector}},
  \href{https://doi.org/10.1103/PhysRevLett.115.131801}{\emph{Phys. Rev. Lett.}
  {\bfseries 115} (2015) 131801},
  [\href{https://arxiv.org/abs/1506.01081}{{\ttfamily 1506.01081}}].

\bibitem{ATLAS-CONF-2017-024}
{\scshape ATLAS Collaboration} collaboration, \emph{{Search for new phenomena
  in events with missing transverse momentum and a Higgs boson decaying into
  two photons at $\sqrt{s}$ = 13 TeV with the ATLAS detector}},  Tech. Rep.
  ATLAS-CONF-2017-024, CERN, Geneva, Apr, 2017.

\bibitem{Bourhis:1997yu}
L.~Bourhis, M.~Fontannaz and J.~P. Guillet, \emph{{Quarks and gluon
  fragmentation functions into photons}},
  \href{https://doi.org/10.1007/s100520050158}{\emph{Eur. Phys. J.} {\bfseries
  C2} (1998) 529--537}, [\href{https://arxiv.org/abs/hep-ph/9704447}{{\ttfamily
  hep-ph/9704447}}].

\bibitem{Bourhis:2000gs}
L.~Bourhis, M.~Fontannaz, J.~P. Guillet and M.~Werlen, \emph{{Next-to-leading
  order determination of fragmentation functions}},
  \href{https://doi.org/10.1007/s100520100579}{\emph{Eur. Phys. J.} {\bfseries
  C19} (2001) 89--98}, [\href{https://arxiv.org/abs/hep-ph/0009101}{{\ttfamily
  hep-ph/0009101}}].

\bibitem{Kniehl:2000fe}
B.~A. Kniehl, G.~Kramer and B.~Potter, \emph{{Fragmentation functions for
  pions, kaons, and protons at next-to-leading order}},
  \href{https://doi.org/10.1016/S0550-3213(00)00303-5}{\emph{Nucl. Phys.}
  {\bfseries B582} (2000) 514--536},
  [\href{https://arxiv.org/abs/hep-ph/0010289}{{\ttfamily hep-ph/0010289}}].

\bibitem{Binnewies:1995kg}
J.~Binnewies, B.~A. Kniehl and G.~Kramer, \emph{{Neutral kaon production in
  $e^{+} e^{-}$, $e p$ and $p \bar{p}$ collisions at next-to-leading order}},
  \href{https://doi.org/10.1103/PhysRevD.53.3573}{\emph{Phys. Rev.} {\bfseries
  D53} (1996) 3573--3581},
  [\href{https://arxiv.org/abs/hep-ph/9506437}{{\ttfamily hep-ph/9506437}}].

\bibitem{Kretzer:2000yf}
S.~Kretzer, \emph{{Fragmentation functions from flavor inclusive and flavor
  tagged e+ e- annihilations}},
  \href{https://doi.org/10.1103/PhysRevD.62.054001}{\emph{Phys. Rev.}
  {\bfseries D62} (2000) 054001},
  [\href{https://arxiv.org/abs/hep-ph/0003177}{{\ttfamily hep-ph/0003177}}].

\bibitem{CMS-PAS-EXO-16-050}
{\scshape CMS Collaboration} collaboration, \emph{{Search for associated
  production of dark matter with a Higgs boson that decays to a pair of bottom
  quarks}},  Tech. Rep. CMS-PAS-EXO-16-050, CERN, Geneva, 2018.

\bibitem{Sirunyan:2018gdw}
{\scshape CMS} collaboration, A.~M. Sirunyan et~al., \emph{{Search for dark
  matter produced in association with a Higgs boson decaying to a pair of
  bottom quarks in proton–proton collisions at $\sqrt{s}=13\,\text {Te}\text
  {V} $}}, \href{https://doi.org/10.1140/epjc/s10052-019-6730-7}{\emph{Eur.
  Phys. J.} {\bfseries C79} (2019) 280},
  [\href{https://arxiv.org/abs/1811.06562}{{\ttfamily 1811.06562}}].

\bibitem{ATLAS-CONF-2016-019}
\emph{{Search for Dark Matter in association with a Higgs boson decaying to
  $b$-quarks in $pp$ collisions at $\sqrt{s} = 13$ TeV with the ATLAS
  detector}},  Tech. Rep. ATLAS-CONF-2016-019, CERN, Geneva, Mar, 2016.

\bibitem{Aaboud:2017yqz}
{\scshape ATLAS} collaboration, M.~Aaboud et~al., \emph{{Search for Dark Matter
  Produced in Association with a Higgs Boson Decaying to $b\bar b$ using 36
  fb$^{-1}$ of $pp$ collisions at $\sqrt s=13$ TeV with the ATLAS Detector}},
  \href{https://doi.org/10.1103/PhysRevLett.119.181804}{\emph{Phys. Rev. Lett.}
  {\bfseries 119} (2017) 181804},
  [\href{https://arxiv.org/abs/1707.01302}{{\ttfamily 1707.01302}}].

\bibitem{ATLAS:2018bvd}
{\scshape ATLAS} collaboration, T.~A. collaboration, \emph{{Search for Dark
  Matter Produced in Association with a Higgs Boson decaying to $b\bar{b}$ at
  $\sqrt{s}= 13\,$TeV with the ATLAS Detector using 79.8$\,$fb$^{-1}$ of
  proton-proton collision data}}, .

\bibitem{Alwall:2014hca}
J.~Alwall, R.~Frederix, S.~Frixione, V.~Hirschi, F.~Maltoni, O.~Mattelaer
  et~al., \emph{{The automated computation of tree-level and next-to-leading
  order differential cross sections, and their matching to parton shower
  simulations}}, \href{https://doi.org/10.1007/JHEP07(2014)079}{\emph{JHEP}
  {\bfseries 07} (2014) 079},
  [\href{https://arxiv.org/abs/1405.0301}{{\ttfamily 1405.0301}}].

\bibitem{Sjostrand:2006za}
T.~Sjostrand, S.~Mrenna and P.~Z. Skands, \emph{{PYTHIA 6.4 Physics and
  Manual}}, \href{https://doi.org/10.1088/1126-6708/2006/05/026}{\emph{JHEP}
  {\bfseries 05} (2006) 026},
  [\href{https://arxiv.org/abs/hep-ph/0603175}{{\ttfamily hep-ph/0603175}}].

\bibitem{deFavereau:2013fsa}
{\scshape DELPHES 3} collaboration, J.~de~Favereau, C.~Delaere, P.~Demin,
  A.~Giammanco, V.~Lemaître, A.~Mertens et~al., \emph{{DELPHES 3, A modular
  framework for fast simulation of a generic collider experiment}},
  \href{https://doi.org/10.1007/JHEP02(2014)057}{\emph{JHEP} {\bfseries 02}
  (2014) 057}, [\href{https://arxiv.org/abs/1307.6346}{{\ttfamily 1307.6346}}].

\bibitem{Cacciari:2006sm}
M.~Cacciari, \emph{{FastJet: A Code for fast $k_t$ clustering, and more}},  in
  \emph{{Deep inelastic scattering. Proceedings, 14th International Workshop,
  DIS 2006, Tsukuba, Japan, April 20-24, 2006}}, pp.~487--490, 2006,
  \href{https://arxiv.org/abs/hep-ph/0607071}{{\ttfamily hep-ph/0607071}}.

\bibitem{Teodorescu:2008zzb}
L.~Teodorescu, \emph{{Artificial neural networks in high-energy physics}},  in
  \emph{{Computing. Proceedings, inverted CERN School of Computing, ICSC2005
  and ICSC2006, Geneva, Switzerland, February 23-25, 2005, and March 6-8,
  2006}}, pp.~13--22, 2008,
  \href{http://doc.cern.ch/yellowrep/2008/2008-002/p13.pdf}{http://doc.cern.ch/yellowrep/2008/2008-002/p13.pdf}.

\bibitem{Roe:2004na}
B.~P. Roe, H.-J. Yang, J.~Zhu, Y.~Liu, I.~Stancu and G.~McGregor,
  \emph{{Boosted decision trees, an alternative to artificial neural
  networks}}, \href{https://doi.org/10.1016/j.nima.2004.12.018}{\emph{Nucl.
  Instrum. Meth.} {\bfseries A543} (2005) 577--584},
  [\href{https://arxiv.org/abs/physics/0408124}{{\ttfamily physics/0408124}}].

\bibitem{Baldi:2014kfa}
P.~Baldi, P.~Sadowski and D.~Whiteson, \emph{{Searching for Exotic Particles in
  High-Energy Physics with Deep Learning}},
  \href{https://doi.org/10.1038/ncomms5308}{\emph{Nature Commun.} {\bfseries 5}
  (2014) 4308}, [\href{https://arxiv.org/abs/1402.4735}{{\ttfamily
  1402.4735}}].

\bibitem{Ghosh:2018gyw}
S.~Ghosh, A.~Harilal, A.~Sahasransu, R.~Singh and S.~Bhattacharya, \emph{{A
  simulation study to distinguish prompt photon from $\pi^0$ and beam halo in a
  granular calorimeter using deep networks}},
  \href{https://doi.org/10.1088/1748-0221/14/01/P01011}{\emph{JINST} {\bfseries
  14} (2019) P01011}, [\href{https://arxiv.org/abs/1808.03987}{{\ttfamily
  1808.03987}}].

\bibitem{Woodruff:2017geg}
{\scshape MicroBooNE} collaboration, K.~Woodruff, \emph{{Automated Proton Track
  Identification in MicroBooNE Using Gradient Boosted Decision Trees}},  in
  \emph{{Proceedings, Meeting of the APS Division of Particles and Fields (DPF
  2017): Fermilab, Batavia, Illinois, USA, July 31 - August 4, 2017}}, 2018,
  \href{https://arxiv.org/abs/1710.00898}{{\ttfamily 1710.00898}},
  \href{http://lss.fnal.gov/archive/2017/conf/fermilab-conf-17-440-e.pdf}{http://lss.fnal.gov/archive/2017/conf/fermilab-conf-17-440-e.pdf}.

\bibitem{Oyulmaz:2019jqr}
K.~Y. Oyulmaz, A.~Senol, H.~Denizli and O.~Cakir, \emph{{Top quark anomalous
  FCNC production via $tqg$ couplings at FCC-hh}},
  \href{https://doi.org/10.1103/PhysRevD.99.115023}{\emph{Phys. Rev.}
  {\bfseries D99} (2019) 115023},
  [\href{https://arxiv.org/abs/1902.03037}{{\ttfamily 1902.03037}}].

\bibitem{Bhattacherjee:2019fpt}
B.~Bhattacherjee, S.~Mukherjee and R.~Sengupta, \emph{{Study of energy
  deposition patterns in hadron calorimeter for prompt and displaced jets using
  convolutional neural network}},
  \href{https://doi.org/10.1007/JHEP11(2019)156}{\emph{JHEP} {\bfseries 11}
  (2019) 156}, [\href{https://arxiv.org/abs/1904.04811}{{\ttfamily
  1904.04811}}].

\bibitem{Hultqvist:1995ibm}
K.~Hultqvist, R.~Jacobsson and K.~E. Johansson, \emph{{Using a neural network
  in the search for the Higgs boson}}, .

\bibitem{Field:1996rw}
R.~D. Field, Y.~Kanev, M.~Tayebnejad and P.~A. Griffin, \emph{{Using neural
  networks to enhance the Higgs boson signal at hadron colliders}},
  \href{https://doi.org/10.1103/PhysRevD.53.2296}{\emph{Phys. Rev.} {\bfseries
  D53} (1996) 2296--2308}.

\bibitem{Bakhet:2015uca}
N.~Bakhet, M.~{\relax Yu}. Khlopov and T.~Hussein, \emph{{Neural Networks
  Search for Charged Higgs Boson of Two Doublet Higgs Model at the Hadrons
  Colliders}},  \href{https://arxiv.org/abs/1507.06547}{{\ttfamily
  1507.06547}}.

\bibitem{Lasocha:2020ctd}
K.~Lasocha, E.~Richter-Was, M.~Sadowski and Z.~Was, \emph{{Deep Neural Network
  application: Higgs boson CP state mixing angle in H to tau tau decay and at
  LHC}},  \href{https://arxiv.org/abs/2001.00455}{{\ttfamily 2001.00455}}.

\bibitem{keras}
J.~R. Hermans, \emph{{Distributed Keras: Distributed Deep Learning with Apache
  Spark and Keras, CERN IT-DB}}, .

\bibitem{tensorflow2015}
\emph{{TensorFlow}},
  \href{https://www.tensorflow.org/}{https://www.tensorflow.org/}.

\bibitem{Hocker:2007ht}
A.~Hocker et~al., \emph{{TMVA - Toolkit for Multivariate Data Analysis}},
  \href{https://arxiv.org/abs/physics/0703039}{{\ttfamily physics/0703039}}.

\bibitem{Kingma:2014vow}
D.~P. Kingma and J.~Ba, \emph{{Adam: A Method for Stochastic Optimization}},
  \href{https://arxiv.org/abs/1412.6980}{{\ttfamily 1412.6980}}.

\bibitem{Chatrchyan:2014goa}
{\scshape CMS} collaboration, S.~Chatrchyan et~al., \emph{{Search for
  supersymmetry with razor variables in pp collisions at $\sqrt{s}$=7 TeV}},
  \href{https://doi.org/10.1103/PhysRevD.90.112001}{\emph{Phys. Rev.}
  {\bfseries D90} (2014) 112001},
  [\href{https://arxiv.org/abs/1405.3961}{{\ttfamily 1405.3961}}].

\bibitem{Fabiani:2020voj}
V.~Fabiani, \emph{{The flavour of Dark Matter. A search for Dark Matter in
  association with a Higgsboson decaying to bottom quarks with the ATLAS
  detector}}, Ph.D. thesis, Nijmegen U., 2020.

\end{thebibliography}\endgroup

\end{document}